# Genome packaging within icosahedral capsids and large-scale segmentation in viral genomic sequences

V. R. Chechetkin[a,b]* and V.V. Lobzin[c]

[a]*Engelhardt Institute of Molecular Biology of Russian Academy of Sciences, Vavilov str., 32, Moscow 119334, Russia*

[b]*Theoretical Department of Division for Perspective Investigations, Troitsk Institute of Innovation and Thermonuclear Investigations (TRINITI), Moscow, Troitsk District 108840, Russia*

[c]*School of Physics, University of Sydney, Sydney, NSW 2006, Australia*

___________________________

*Corresponding author. *E-mail addresses:* chechet@eimb.ru; vladimir_chechet@mail.ru

Tel.: +7 499 135 9895. Fax: +7 499 135 1405 (V.R. Chechetkin).

**Abstract**

The assembly and maturation of viruses with icosahedral capsids must be coordinated with icosahedral symmetry. The icosahedral symmetry imposes also the restrictions on the cooperative specific interactions between genomic RNA/DNA and coat proteins that should be reflected in quasi-regular segmentation of viral genomic sequences. Combining discrete direct and double Fourier transforms, we studied the quasi-regular large-scale segmentation in genomic sequences of different ssRNA, ssDNA, and dsDNA viruses. The particular representatives included satellite tobacco mosaic virus and the strains of satellite tobacco necrosis virus STNV-C, STNV-1, STNV-2, Escherichia phages MS2, $\phi$X174, α3, and HK97, and Simian virus 40. In all their genomes, we found the significant quasi-regular segmentation of genomic sequences related to the virion assembly and the genome packaging within icosahedral capsid. We also found good correspondence between our results and available cryo-electron microscopy data on capsid structures and genome packaging in these viruses. Fourier analysis of genomic sequences provides the additional insight into mechanisms of hierarchical genome packaging and may be used for verification of the concepts of 3-fold or 5-fold intermediates in virion assembly. The results of sequence analysis should be taken into account at the choice of models and data interpretation. They also may be helpful for the development of antiviral drugs.

**Keywords**

**List of Abbreviations**

AS, assembly signals; DDFT, discrete double Fourier transform; DFT, discrete Fourier transform; dsDNA, double-stranded DNA; ORF, open reading frame; PS, packaging signals;

ssDNA, single-stranded DNA; ssRNA, single-stranded RNA; STMV, satellite tobacco mosaic virus; STNV, satellite tobacco necrosis virus; SV40, Simian virus 40; UTR, untranslated region

# 1. Introduction

In the virus world, more than a half of viruses belongs to the viruses with spherical icosahedral capsids (Calendar & Abedon, 2006; Rossmann & Rao, 2012; Mateu, 2013). The icosahedral symmetry determines the pathways of virion assembly and maturation of these viruses. The genome packaging can be performed alongside with capsid assembly and may actively affect the virion assembly (as in ssRNA viruses), whereas in the other cases the packaging is performed into pre-synthesized capsid with special molecular machinery (as in dsDNA bacteriophages) (Roos et al., 2007; Sun et al., 2010; Aksyuk & Rossmann, 2011; Black & Thomas, 2012; Cuervo et al., 2013). The theory by Caspar & Klug (1962) established a basis for classification of virus capsids in terms of triangulation numbers, $T$. The completion of icosahedral symmetry by affine transforms (Twarock, 2006) permits to describe nearly all observable capsid structures. A similar theory for genome packaging within icosahedral capsids is absent and even the constructive theoretical approach to this problem is not yet explicitly formulated. The experimental structural data based on the X-ray crystallography, cryo-electron microscopy (cryo-EM) and cryo-electron tomography, atomic force microscopy et al. (Mateu, 2013) together with molecular dynamic simulations remain the main source of our knowledge on the genome packaging within icosahedral capsids.

The cooperative specific interactions between genomic DNA or RNA and capsomers and the general restrictions imposed by the icosahedral symmetry on the genome packaging within icosahedral capsids should yield the large-scale segmentation in genomic nucleotide sequences. Due to symmetry such large-scale segmentation ought to be quasi-periodic and can be detected by Fourier analysis of underlying genomic sequences (for a review and further references see, e.g., Lobzin & Chechetkin, 2000). With few exceptions, the underlying quasi-periodic patterns are fuzzy due to the multiple point mutations and insertions/deletions particularly important for rapidly evolving viral genomes. The related quasi-periodic patterns are superimposed with the other patterns participating in the different molecular mechanisms during virus life cycle.

Therefore, their detection needs sophisticated statistical methods with multiple cross-check. Combining discrete direct and double Fourier (Chechetkin & Lobzin, 2017) transforms, we studied the large-scale quasi-periodic patterns in the genomic sequences of satellite tobacco mosaic virus (STMV) and the strains of satellite tobacco necrosis virus (STNV), STNV-1, STNV-2, STNV-C, Escherichia virus MS2 (all are positive-strand ssRNA viruses), Escherichia viruses ϕX174 and α3 (ssDNA viruses), Macaca mulatta polyomavirus 1 (or Simian virus 40, SV40, with circular dsDNA minichromosome genome), and Escherichia virus HK97 (with linear dsDNA genome). In all these genomes, the combined Fourier analysis revealed statistically significant quasi-periodic patterns associated with the elements of icosahedral symmetry. A part of quasi-periodic patterns appears to be reproducible and inherent to all genomes, whereas the other part is specific to the particular genome. For ssRNA viruses some of detected segmentation modes are directly related to the two-stage packaging mechanism and packaging signals suggested by Stockley et al. (2016). We will show that the assembly and packaging signals are different. For ssDNA viruses a part of modes can be associated with the transport of ssDNA bound with DNA-binding proteins J to procapsid and subsequent packaging, whereas for SV40 we found the set of periodicities related to the dynamic positioning of nucleosomes on dsDNA. The characteristic segmentation modes for HK97 are grouped within range between persistence and Kuhn lengths. The detected patterns proved to be in good correspondence with the experimental data (where available) on the capsid structures and the genome packaging.

The results of sequence analysis can be used for the adjustment of experimental data and models of genome packaging. The detection of three-segment or five-segment patterns in the genome may be used for the verification of the concepts of 3-fold or 5-fold intermediates in virion assembly. Generally, the combined Fourier analysis yields additional insight into hierarchical organization of viral genomes related to the basic mechanisms of virus functioning. The study of these mechanisms provides molecular ground for the development of antiviral drugs and targeted therapy.

## 2. Theory and methods

### *2.1. Icosahedral symmetry*

The icosahedral symmetry comprises 15 axes of the second order, 10 axes of the third order, and 6 axes of the fifth order. The total number of operations for the icosahedral symmetry is 60. The global icosahedral symmetry of viral capsids can be characterized either in terms of the axes (like in the RCSB PDB; URL rcsb.org; Berman et al., 2000) or in terms of dual icosahedron/dodecahedron representation (like in the VIPERdb; URL viperdb.scripps.edu; Carrillo-Tripp et al., 2009). In the latter case, the corresponding perfect solid figures have 30 edges, 20 faces/vertices, and 12 vertices/faces. The correspondence should be searched between the elements of icosahedral symmetry (characterized in terms of both representations) and the numbers of large-scale quasi-periodic patterns in genomic sequences. We also tried the extended version of such correspondence with the multiple numbers of symmetry elements.

### *2.2. Discrete direct and double Fourier transforms*

In this section we follow the methods developed previously (Chechetkin & Turygin, 1994, 1995; Lobzin & Chechetkin, 2000). In discrete Fourier transform (DFT) harmonics corresponding to nucleotides of type $\alpha \in (A, C, G, T)$ in a genomic sequence of length $M$ are calculated as

$$\rho_\alpha(q_n) = M^{-1/2} \sum_{m=1}^{M} \rho_{m,\alpha} e^{-iq_n m}, \quad q_n = 2\pi n/M, \quad n = 0, 1, ..., M-1 \qquad (1)$$

Here $\rho_{m,\alpha}$ indicates the position occupied by the nucleotide of type $\alpha$; $\rho_{m,\alpha} = 1$ if the nucleotide of type $\alpha$ occupies the $m$-th site and 0 otherwise. The amplitudes of Fourier harmonics (or structure factors) are defined as

$$F_{\alpha\alpha}(q_n) = \rho_\alpha(q_n)\rho_\alpha^*(q_n) \qquad (2)$$

where the asterisk denotes the complex conjugation. The harmonics with $n = 0$ depend only on the nucleotide composition. They do not contain structural information and will be discarded. Due to the symmetry of structure factors,

$$F_{\alpha\alpha}(q_n) = F_{\alpha\alpha}(2\pi - q_n) \qquad (3)$$

Fourier spectrum can be restricted from $n = 1$ to

$$N = [M/2] \qquad (4)$$

where the brackets denote the integer part of the quotient. The structure factors will always be normalized relative to the mean spectral values, which are determined by the exact sum rules,

$$f_{\alpha\alpha}(q_n) = F_{\alpha\alpha}(q_n)/\overline{F}_{\alpha\alpha}; \quad \overline{F}_{\alpha\alpha} = N_\alpha(M - N_\alpha)/M(M-1) \qquad (5)$$

where $N_\alpha$ is the total number of nucleotides of type $\alpha$ in a sequence of length $M$. Below, we will use the sums

$$S_{AT}(q_n) = f_{AA}(q_n) + f_{TT}(q_n); \, S_{GC}(q_n) = f_{GG}(q_n) + f_{CC}(q_n) \qquad (6)$$

$$S_4(q_n) = f_{AA}(q_n) + f_{CC}(q_n) + f_{GG}(q_n) + f_{TT}(q_n) \qquad (7)$$

which remain invariant under the mutual replacements between complementary nucleotides. These combinations are convenient in the study of stem-loop units in ssRNA genomes or in the study of genomes with gene coding on both strands in dsDNA. There is direct correspondence between the number of quasi-periodic patterns, $N_s$, and the spectral number $n$ in the sums (6)–(7),

$$N_s = n \qquad (8)$$

The period $p$ is measured in terms of the number of nucleotides or base pairs (these units will always be tacitly implied below) and is calculated as,

$$p = M/n \qquad (9)$$

The periodic patterns generate a series of equidistant peaks in the space of spectral numbers $n$ (Chechetkin & Turygin, 1995; Lobzin & Chechetkin, 2000; Sharma et al., 2004; Paar et al., 2008). For a period $p = M/n$, the corresponding series comprises the numbers $n$, $2n$, ..., $k_{max}n \leq N$. The random variations in periods of patterns induce the deviations in equidistance and may partially suppress the harmonics with the higher numbers $kn$ (Lobzin & Chechetkin, 2000). Generally, the periodic patterns randomized by the point mutations and/or indels can be detected either by statistically significant singular high peaks and/or by the sums of equidistant harmonics (Chechetkin & Turygin, 1995; Chechetkin & Lobzin, 1998; Lobzin & Chechetkin, 2000), the feature related to equidistant series being unique and more important statistically.

The large-scale periodic patterns generate the equidistant series long enough to be detected by the iteration of Fourier transform or by discrete double Fourier transform (DDFT) (Chechetkin & Lobzin, 2017). DDFT provides the efficient tool for filtering false positives and false negatives in the primary DFT spectra. The false positives in DFT spectra can be produced by the random outbursts or by the effects unrelated to periodic patterns (e.g., by the mosaic patchiness of the genome or by the non-periodic large-scale variations in the nucleotide composition), whereas the false negatives correspond to the missed patterns which can be detected by the equidistant series.

The harmonics in DDFT are calculated as

$$\Phi(\tilde{q}_{n'}) = (N-1)^{-1/2} \sum_{n=2}^{N} S(q_n) e^{-i\tilde{q}_{n'}n}, \quad \tilde{q}_{n'} = 2\pi n'/(N-1), \quad n' = 0, 1, ..., N-2 \quad (10)$$

where $N$ is defined by Eq. (4) and the sums $S(q_n)$ are defined by Eqs. (6) and (7). The DFT harmonic with $n = 1$ does not induce equidistant series and is discarded from DDFT. The other definitions for DDFT are in lines with that for DFT. The amplitudes of harmonics (10) are given by

$$F_{II}(\tilde{q}_{n'}) = \Phi(\tilde{q}_{n'})\Phi^*(\tilde{q}_{n'}) \qquad (11)$$

Again, we are interested in harmonics with non-zero spectral numbers $n'$ and can restrict the spectrum to the left half from $n' = 1$ to

$$N' = [(N-1)/2] \qquad (12)$$

due to the symmetry relationship similar to Eq. (3). The amplitudes (11) are normalized as

$$f_{II}(\tilde{q}_{n'}) = F_{II}(\tilde{q}_{n'})/\overline{F}_{II} \qquad (13)$$

$$\overline{F}_{II} = \frac{1}{N'}\sum_{n'=1}^{N'} F_{II}(\tilde{q}_{n'}) \qquad (14)$$

Generally, the equidistant series in DFT spectra also generate the corresponding equidistant series in DDFT spectra with spectral numbers $k'n'$, $k' = 1, ..., k'_{max}$; $k'_{max} n' \leq N'$, where $N'$ is defined by Eq. (12)

The number of quasi-periodic segments can be assessed by the spectral number $n'$ for the singular high amplitude $f_{II}(\tilde{q}_{n'})$ as

$$N'_s = (N-1)/n' \qquad (15)$$

whereas their periods in nucleotides or base pairs are given by

$$p'_s = M/N'_s \qquad (16)$$

The strict or approximate correspondence between the number of quasi-periodic segments determined by DFT and DDFT,

$$N_s \approx N'_s \qquad (17)$$

strongly enhances their statistical significance. Thus, the combination of DFT and DDFT provides the necessary cross-check for the detection of fuzzy repeating patterns. DDFT resolves the equidistant DFT series with spectral numbers satisfying the inequality

$$\sqrt{N} > n \qquad (18)$$

where *n* corresponds to the spectral number associated with the large-scale periodicity under study. The inequality (18) restricts the left side of DDFT spectra by a boundary about $\sqrt{N}$. In the subsequent analysis we will extrapolate the spectral range in DDFT partly beyond the boundary corresponding to (18).

### *2.3. Statistical criteria and preprocessing of spectra*

The probability that the sums (6)–(7) for DFT of random sequences exceed a fixed value *S′* is given by (Lobzin & Chechetkin, 2000),

$$\Pr(S > S') = e^{-S'} \sum_{k=1}^{r} \frac{S'^{k-1}}{(k-1)!}; \; r = 2, 4 \qquad (19)$$

whereas the probability that a normalized harmonic (13) for DDFT exceeds a fixed value *f′* obeys Rayleigh distribution (Chechetkin & Lobzin, 2017),

$$\Pr(\tilde{f}_{II} > f') = e^{-f'} \qquad (20)$$

In this paper we will apply the lowest acceptable statistical threshold for the probabilities (19) and (20), Pr = 0.05, supplemented by Pr = 0.01 for the comparison and ranking. We found that criteria based on the rigorous extreme value statistics are too strict for the problem concerned.

The statistical assessment needs that the spectral ranges did not display any visible trends. In particular, the highest harmonics in DFT spectra (commonly related to the coding periodicity *p* = 3) can generate broad trends in DDFT spectra comparable to the whole spectrum. To suppress such trends, we imposed the thresholds on the sums for DFT spectra based on the extreme value probability in Bonferroni approximation, Pr = 0.05/$N$. If the probability for the sums (6)–(7) appears to be lower than this value, the heights of the sums are restricted by the threshold corresponding to Pr = 0.05/$N$. After such cut-off, the trends in DDFT spectra become narrow and can be leveled by the methods developed earlier (Chechetkin & Lobzin, 2017).

Due to the restriction (18) the ranges of DDFT under study do not comprise the shorter quasi-periodic patterns, which can be studied separately by the modification of the main scheme (Chechetkin & Lobzin, 2017). We preferred to retain the unified approach and used the following procedure for the assessment of the shorter quasi-periodic patterns by DDFT analysis. We picked up the harmonics in equidistant DDFT series with a window $\delta$

$$n'_I = k'\{n'_s\} \pm \delta; \; k' = k'_{min}, ..., k'_{max} \qquad (21)$$

$$n'_{II} = \{k'n'_s\} \pm \delta; \; k' = k'_{min}, ..., k'_{max} \qquad (22)$$

for which the corresponding probability (20) is lower than 0.05. Here the braces mean the rounding to the nearest integer. Because of the ambiguity in the rounding procedure we used both modes of rounding. The spectral number $n'_s$ is related to the number of quasi-periodic patterns by reciprocating Eq. (15) and $k'_{min}, k'_{max}$ are determined by the boundaries of the chosen DDFT range, $k'_{min}n'_s \geq n'_{min}; k'_{max}n'_s \leq N'$. The integers $k'$ vary within the interval $(k'_{min}, k'_{max})$. For the random sequences the mean number of such picked up harmonics would be

$$\overline{N}_{random} = 0.05 N_w \qquad (23)$$

where $N_w$ is the total number of harmonics in the corresponding equidistant series with window $\delta$. If the overlapping between windows and the boundaries of the spectral range is absent, their number is $N_w = (k'_{max} - k'_{min} + 1)(2\delta + 1)$. The significance of $N_{obs}$ observed harmonics can be assessed by Poisson probability

$$\Pr = \frac{(\overline{N}_{random})^{N_{obs}}}{N_{obs}!} \exp(-\overline{N}_{random}) \qquad (24)$$

The similar scheme can be applied to the assessment of enrichment of particular spectral ranges by the higher harmonics. The intricate network of commensurate periods may be considered as a characteristic feature of genetic sequences differing them from the random ones. The differentiation between real (determined by the actual genetic mechanisms) and formal (determined by the Fourier transform formalism) peaks is not strict for the fuzzy repeats and needs additional study. As we are primarily interested in the detection of the regular features in underlying genomic sequences, the mere detection of significant commensurate peaks can be used as a distinguishing sign of regularity.

The search for the significant commensurate harmonics can serve as a means of additional cross-check. To sum up, any DFT harmonic exceeding 5%-significance threshold and potentially related to the elements of icosahedral symmetry is tried by the multiple cross-check scheme including: (i) search for the significant counterpart harmonics for nucleotides of different types; (ii) search for the significant commensurate harmonics; (iii) search for the clustering of significant harmonics around periodicity under study; (iv) combined application of DFT and DDFT and search for the correspondence between harmonics in two transforms (the items i–iii concern the significant DDFT harmonics as well); (v) application of the test with equidistant series in DDFT spectra. Including 5%-significance threshold, the filtering by all six conditions is excessively strong. As the probabilities for the separate filters are multiplied, the fulfillment of 3–4 conditions from the list above is quite enough for the practical applications. Such multiple cross-check ensures the reliable detection of relatively weak statistical features.

## *2.4. Reconstruction of motifs corresponding to peak harmonics*

The motifs corresponding to the peak harmonics are of primary interest for potential medical and biological applications. The consensus patterns associated with the most pronounced periodicities in DFT spectra can be reconstructed by the following scheme. The nucleotides of the type α in the sites $m_p = m_0 + \{kp\}$, $m_0 = 1, ..., \{p\}$, $k = 0, ..., N_s - 1$ should be counted and recalculated into the corresponding frequencies (where $p = M/n$, $n = N_s$, the braces denote the rounding to the nearest integer; during counting $m_0$ and $p$ are fixed, $k$ is running). The calculated frequencies $f_{\alpha;p}(m_0)$ can be assessed versus the related frequencies over the genome $f_\alpha$ via the approximate Gaussian criterion,

$$z_{\alpha;p}(m_0) = \left(f_{\alpha;p}(m_0) - f_\alpha\right) / \left(f_\alpha(1 - f_\alpha)/N_s\right)^{1/2} \tag{25}$$

The standard 5% and 1% significance thresholds for the $z$-criterion (25) are ±1.96 and ±2.58. In some cases we will add the nearly significant positions with $0.05 < \Pr < 0.06$ corresponding to $1.89 < |z| < 1.96$. This scheme is applicable both to four-nucleotide representation of genomic sequences and to the binary representations R-Y, S-W, and K-M.

## 3. Results

### 3.1. Large-scale periodic patterns in STMV and STNV genome

STMV, a small icosahedral plant virus with linear positive-strand ssRNA genome, may be considered as one of the smallest reproducing species in nature (for a review and further references see Dodds, 1998; Larson & McPherson, 2001). Its reproducibility needs both the host cell and a host virus (tobacco mosaic virus in this case). The icosahedral capsid consisting of 60 identical subunits ($T = 1$) with genomic ssRNA inside was resolved on 1.4 Å scale (Larson et al., 2014; PDB: 4OQ8; see also Fig. 1A taken from PDB). Electron-density maps provided data on nearly 57% of the genomic ssRNA (the inside view of capsid is taken from VIPERdb and shown in Fig. 1B). The visible RNA revealed 30 double-helical segments, each about 9 bp in length, packaged along the edges of capsid icosahedron (see Fig. 1C taken from the paper by Zeng et al., 2012). Besides these short stem-loop units, the long-range base pairing was also observed in the genome organization of STMV (Archer et al., 2013; Athavale et al., 2013).

The genomic sequence of STMV is of length $M = 1058$ (GenBank: M25782). About a half of the genome contains two overlapping ORF, the longer of which codes for coat protein, whereas the other half contains UTR. The corresponding DFT and DDFT spectra for the genomic sequence of STMV are shown in Fig. 2 (see also the summary of data in Supplementary file S1). All spectra were analyzed, first, in the complete ranges. For the presentation purposes we restricted DFT spectra to the range related to the large-scale segmentation of the genome. The left boundary for DDFT is restricted by inequality (18) (here and below DDFT spectra will be partly extrapolated beyond this boundary).

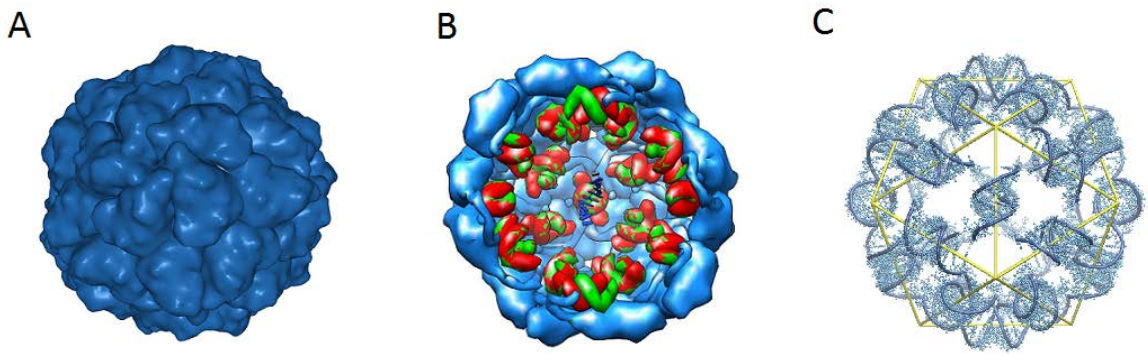

**Figure 1.** The genome packaging within STMV capsid. (A) The surface overview of STMV capsid. (B) The inside view of STMV capsid with the fragments of resolved RNA stems. (C) The scheme of ssRNA genome packaging within icosahedral capsid (Zeng et al., 2012).

The DFT spectrum for the sum $S_{GC}(q_n)$ (Eq. (6); henceforth G+C for brevity) reveals the high peak at $n = 30$ (Pr = $3.94 \times 10^{-6}$) corresponding to 30 stem-loop units in accordance with the data by Larson et al. (2014). The approximate counterpart peak is also seen in DDFT spectrum for $f_{II, S_{GC}}(\tilde{q}_{n'})$ (Eq. (13); we will use the notation G+C for these harmonics as well, adding DDFT for the differentiation from DFT) that proves the "true periodicity" for 30 stem-loop segmentation. The harmonic with $n = 29$ is the third in the ranking for the DFT sum $S_{AT}(q_n)$ (henceforth A+T for brevity) and also can be referred to 30 stem-loop units along icosahedron edges. The highest peak in DFT range for A+T corresponds to the harmonic with $n = 13$ (Pr = $4.15 \times 10^{-4}$) that can be approximately referred to icosahedron vertices. We did not observe the significant association between the quasi-periodic segmentation and icosahedron faces in both DFT and DDFT spectra. The harmonics corresponding to the approximate doubling of 30-element segmentation were detected in DFT spectra for both A+T and G+C ($n = 64$ for A+T and $n = 54, 55$ for G+C; see also $N'_s = 66.0$ in DDFT spectra for G+C and $S_4$). As our definition of DFT is invariant with respect to complementary replacements between nucleotides, such doubling of segmentation can be related to the complementary fragments in the stems. Such doubled segmentation may also be referred to the whole icosahedral symmetry. The quasi-periodic segmentation which can be attributed to the edges, vertices and the whole icosahedral symmetry was also detected both in DDFT spectra (Fig. 2, right) and by the test with equidistant series in DDFT spectra (Section 2.3). The summary of these results is presented in Supplementary files S1 and S2.

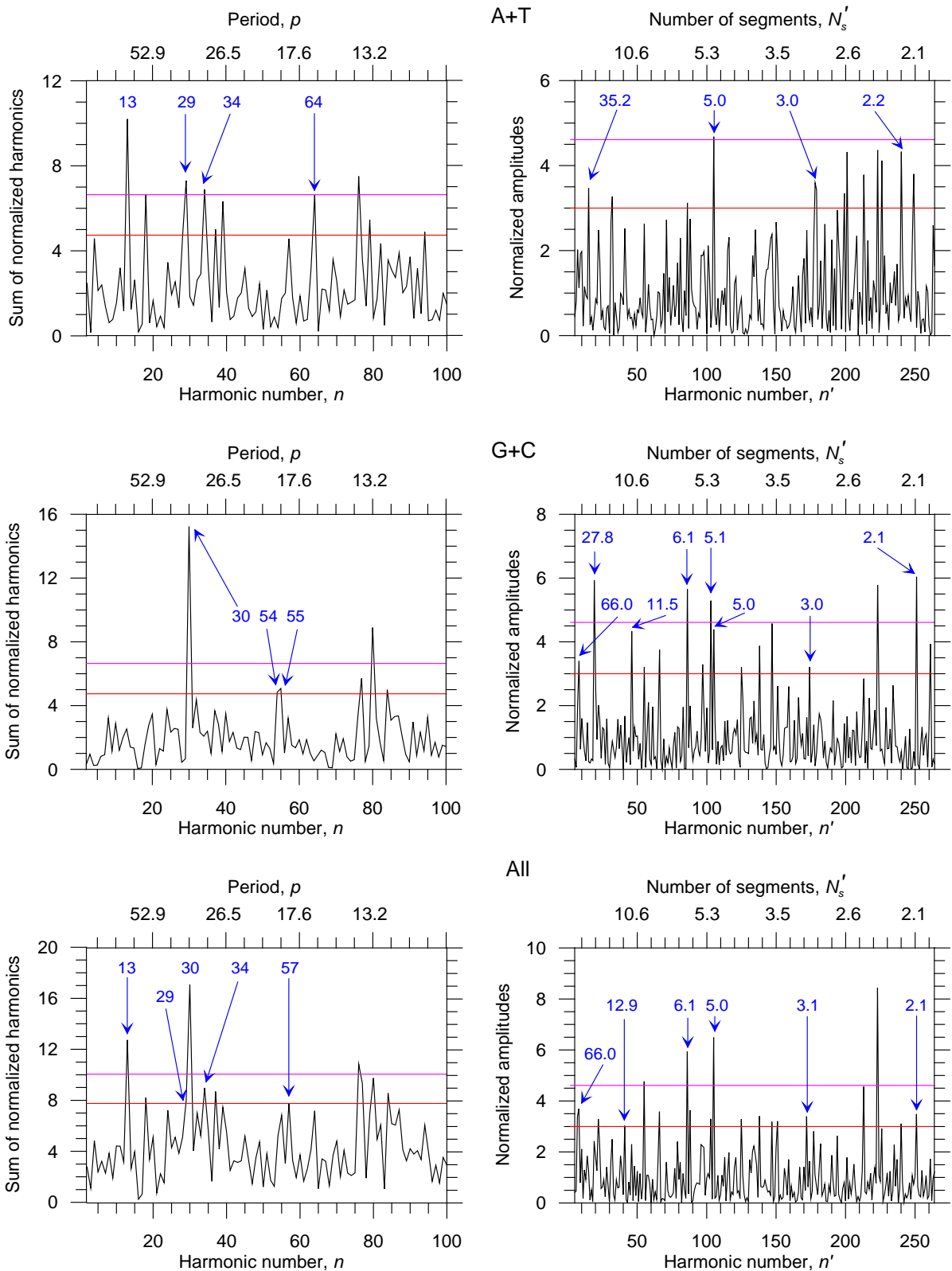

**Figure 2.** The DFT (left) and DDFT (right) spectra for the STMV genomic sequence. The significant harmonics related to the elements of icosahedral symmetry are marked by arrows. The numbers of quasi-periodic segments corresponding to the marked harmonics are shown explicitly in the panels. The red horizontal line corresponds to Pr = 0.05 for a particular harmonic, whereas the magenta horizontal line corresponds to Pr = 0.01.

The RNA folding within icosahedral capsid and in solution is believed to be different (Larson & McPherson, 2001; Athavale et al., 2013), though the experiments indicate the comparable level of RNA compaction in both cases (Kuznetsov et al., 2005). In A+T DFT spectrum we found the additional segmentation corresponding to the harmonics at $n = 34, 37$ (see also the counterpart harmonic in A+T DDFT spectrum with $N'_s = 35.2$) and the approximately doubled segmentation corresponding to the harmonic at $n = 76$ that may be attributed to the difference in RNA folding depending on external conditions.

The capsid assembly is presumed to be performed hierarchically via 3-fold (Sorger et al., 1986) or 5-fold (Rossmann et al., 1983) intermediates. As RNA participates actively in virion assembly, the similar mechanism may be suggested for RNA packaging. Molecular dynamics simulations (Freddolino et al., 2006) also indicate this possibility. DDFT technique was specifically developed for search for large-scale quasi-periodic segmentation in DNA/RNA sequences. The related information is presented in Supplementary file S1 in the columns DDFT spectra. In particular, for STMV the segmentation $n' = 105$, $N'_s \approx 5.03$ corresponds to the highest harmonic in A+T DDFT spectrum (Pr = 0.009). Besides, the corresponding harmonic with $n' = 105$ is approximately commensurate with harmonic $n' = 213$ exceeding the threshold of 5% significance in this spectrum. The harmonic with the same wave number $n' = 105$ was also observed among DDFT harmonics exceeding the threshold of 5% significance in G+C spectrum (Pr = 0.012) and was the second ranked for DDFT $S_4$ spectrum (Pr = 0.0015). The approximately commensurate harmonic with $n' = 213$ was also detected among harmonics exceeding the threshold of 5% significance for DDFT $S_4$ spectrum. The simultaneous observation of these features makes the detection of segmentation $n' = 105$, $N'_s \approx 5.03$ statistically reliable. The significant quasi-periodic segmentation with $N'_s \approx 5$ detected by DDFT spectra agrees with the suggestion on 5-fold intermediate, though multi-channel mechanism with competing 3-fold intermediate cannot be excluded as well.

In addition to the genome of STMV, we also studied the large-scale quasi-periodic patterns in the genomes of the strains STNV-C, STNV-1, and STNV-2 (GenBank: AJ000898, NC_001557, and M64479). Their lengths are distinctly longer in comparison with STMV, $M =$ 1221, 1239, and 1245. The nucleotide content ratio (A+T)/(G+C) is 1.18 (STMV), 1.35 (STNV-C), 1.11 (STNV-1), and 1.09 (STNV-2). The details of DFT and DDFT analysis for STMV and STNV can be found in Supplementary files S1 and S2 on STMV page. Here, we present briefly the main results.

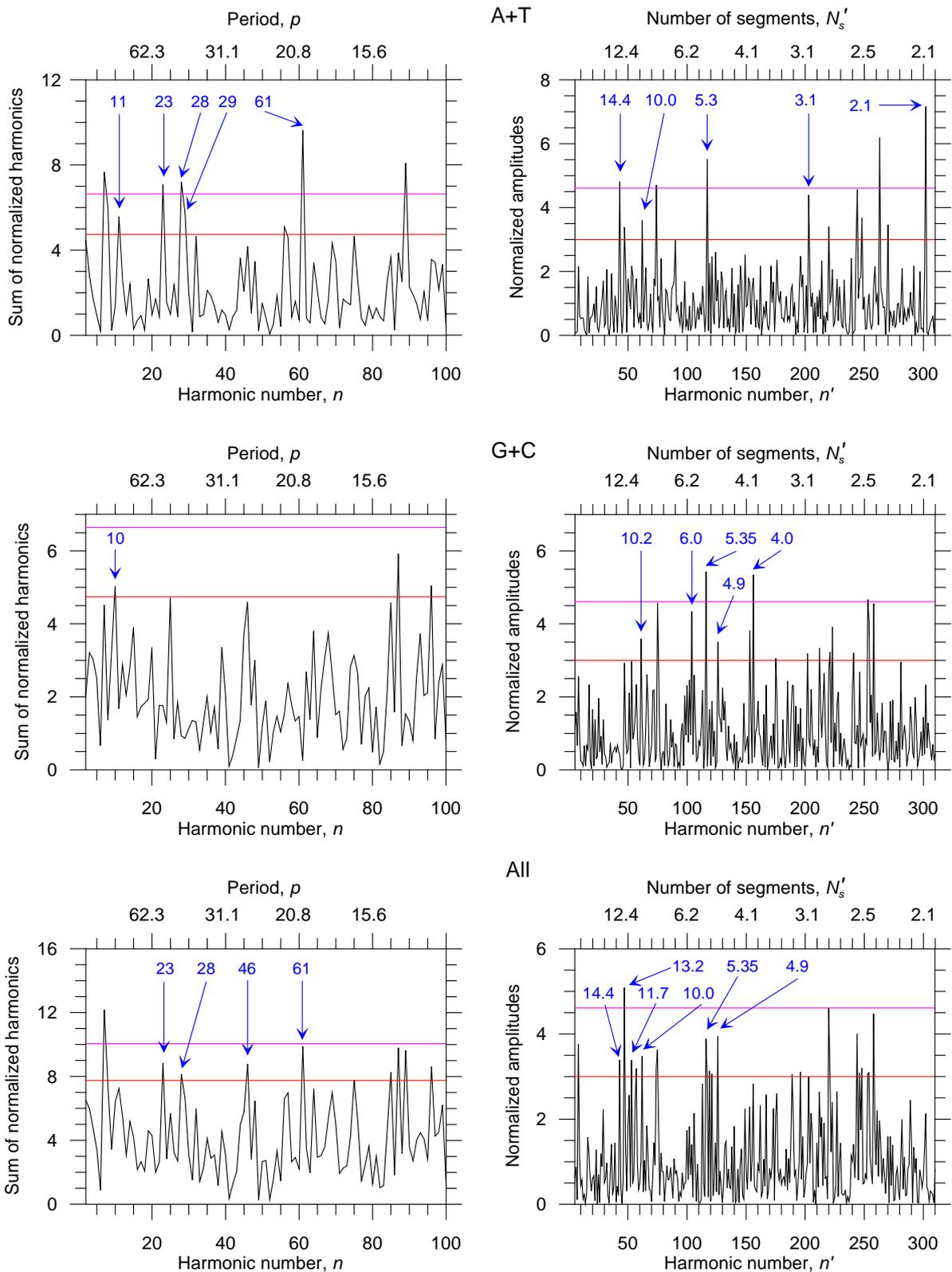

**Figure 3.** The DFT (left) and DDFT (right) spectra for the STNV-2 genomic sequence. The significant harmonics related to the elements of icosahedral symmetry are marked by arrows. The numbers of quasi-periodic segments corresponding to the marked harmonics are shown

explicitly in the panels. The red horizontal line corresponds to Pr = 0.05 for a particular harmonic, whereas the magenta horizontal line corresponds to Pr = 0.01.

The large-scale quasi-periodic patterns for STNV-C appeared to be approximately similar to that for STMV. The most pronounced harmonic in DFT spectra corresponds to $n = 33$, $p = 37.0$ (G+C, Pr = $8.72 \times 10^{-3}$) and is accompanied by the approximately doubled segmentation, $n = 68$, $p = 18.0$ (A+T, G+C, $S_4$). The related harmonics with approximately doubled periods were also observed in DDFT spectra, $n' = 40$, $N'_s = 15.2$, $p' = 80.2$ (A+T, $S_4$). The other characteristic harmonics in DDFT spectra were associated with the icosahedron vertices, $n' = 49$, $N'_s = 12.4$ (A+T); $n' = 52$, $N'_s = 11.7$ (A+T, G+C, $S_4$).

The patterns for STNV-1 and STNV-2 appeared to be strongly different from that for STMV. The related DFT and DDFT spectra for STNV-2 are shown in Fig. 3. The highest harmonic in DFT spectra was observed at $n = 61$, $p = 20.4$ (A+T, Pr = $6.94 \times 10^{-4}$) and was accompanied by the harmonics, $n = 28$, $p = 44.5$; $n = 29$, $p = 42.9$ (A+T) corresponding to the approximately doubled periods. The approximate doubling of the latter periods was observed in DDFT spectra, $n' = 43$, $N'_s = 14.4$, $p' = 86.2$ (A+T, $S_4$). The test with equidistant harmonics in DDFT spectra revealed the significant modes with 12, 20, and 60 segments for STNV-2 (see Supplementary file S2). The patterns for STMV-1 were closer to STNV-2 but turned out to be more distorted and fuzzier. In particular, the highest harmonic in A+T DFT spectrum was at $n = 52$, $p = 23.8$. This harmonic was also the highest in $S_4$ DFT spectrum (Pr = 0.004). The approximately commensurate harmonic at $n = 25$, $p = 49.6$ (Pr = 0.02) was detected in G+C DFT spectrum. Therefore, the spectra for STNV-1 can be considered as intermediate between that for STMV and STMV-2. The characteristic harmonics in DDFT spectra for STNV-1 were associated with the icosahedron faces, $n' = 29$, $N'_s = 21.3$ (the highest harmonic in A+T DDFT spectrum, Pr = $8.96 \times 10^{-4}$, and the second ranked harmonic in $S_4$ DDFT spectrum).

The graphical representation of motif reconstruction scheme for the characteristic peak harmonics is shown in Supplementary figure S3.

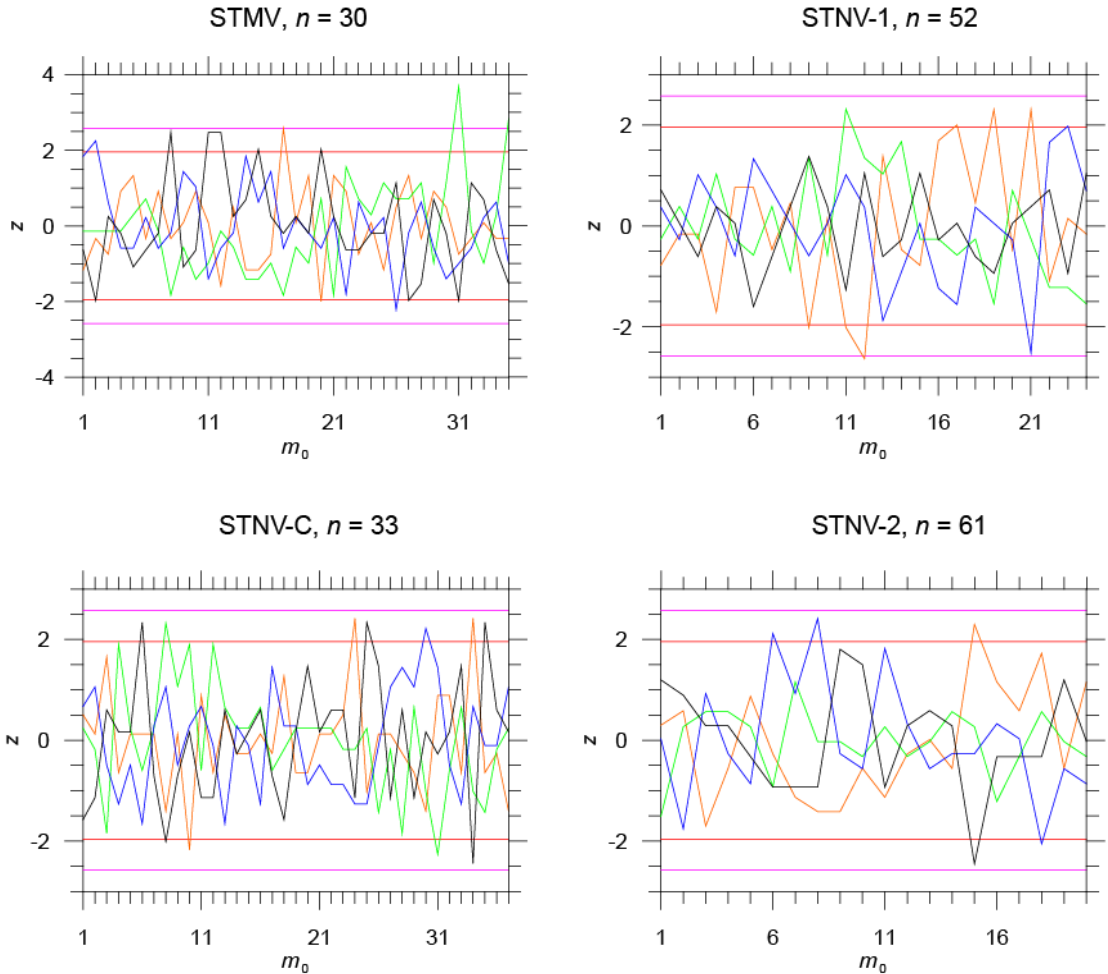

**Figure S3.** The reconstruction of motifs associated with the characteristic peak harmonics in DFT spectra for the genomes of STMV and STNV (Section 2.4). The colors for $z$-profiles (Eq. (25)) corresponding to the nucleotides of different types are: A, orange; G, green; T, blue; and C, black. The red horizontal line corresponds to the significance threshold of $Pr = 0.05$, whereas the magenta horizontal line corresponds to $Pr = 0.01$.

We reconstructed the following patterns for the leading periodicities (Section 2.4): STMV, $n = 30$, $NtN_5cN_2ccN_2cNAN_2cN_{10}GN_3G$; STNV-C, $n = 33$, $N_3gNcNgN_gNgN_{11}acN_4tN_3acN_2$ ($z = 1.90$, $Pr = 0.06$ for $g$); STNV-1, $n = 52$, $N_{10}gN_4a aNaNaNtN$ ($z = 1.69$, $Pr = 0.09$ for $a$); STNV-2, $n = 61$, $N_5tNtN_6aN_2aN_2$ ($z = 1.73$, $Pr = 0.08$ for $a$). In the consensus patterns the lower case letters correspond to the statistical significance within the range 1–5%, the upper case letters correspond to the statistical significance below the threshold of 1%, whereas the non-conservative positions with significance above 5% can be filled by the nucleotides of any type N. In some cases we added the nearly significant nucleotides (the lower case italic letters) to stress the correspondence between the counterpart motifs (cf. $GN_3G$ for STMV and $gNcNgNgNg$ for STNV-C or $a$aNaNa for STNV-1 and $aN_2a$ for STNV-2 as well as the motif $AN_2A$ suggested by Patel et al. (2017)).

The harmonics with the periodicities around $p \approx 11.0$ were also detected in DFT spectra for STMV and STNV. Such periodicities may be attributed to the pitch of A-dsRNA indicating the phasing of stem positions along the secondary RNA structure. The largest segmentation assessed via DDFT spectra in genomic sequences for all strains was associated with $N'_s \approx 2$ and $N'_s \approx 5$.

### *3.2. Large-scale periodic patterns in MS2 genome*

The positive-strand ssRNA Escherichia virus MS2 illustrates how the assembly mechanisms depend on the genome length and capsid structure. Its genomic ssRNA is actively involved in the assembly and maturation of virion (Valegård et al., 1997; Basnak et al., 2010; Rolfsson et al., 2016; Stockley et al., 2016). 180 copies of coat protein assemble to form a *T*=3 icosahedral capsid of MS2. The coat proteins exist in three distinct conformations, A, B, and C, that form dimers A/B and C/C. There are 60 A/B dimers and 30 C/C dimers. The general structure of MS2 capsid is shown in Figs. 4A–4C (taken from VIPERdb; PDB: 1ZDH). One of C/C dimers is replaced by maturation protein responsible for attaching the virus to an F-pilus of *E. coli* and delivering the viral genome into the host during infection.

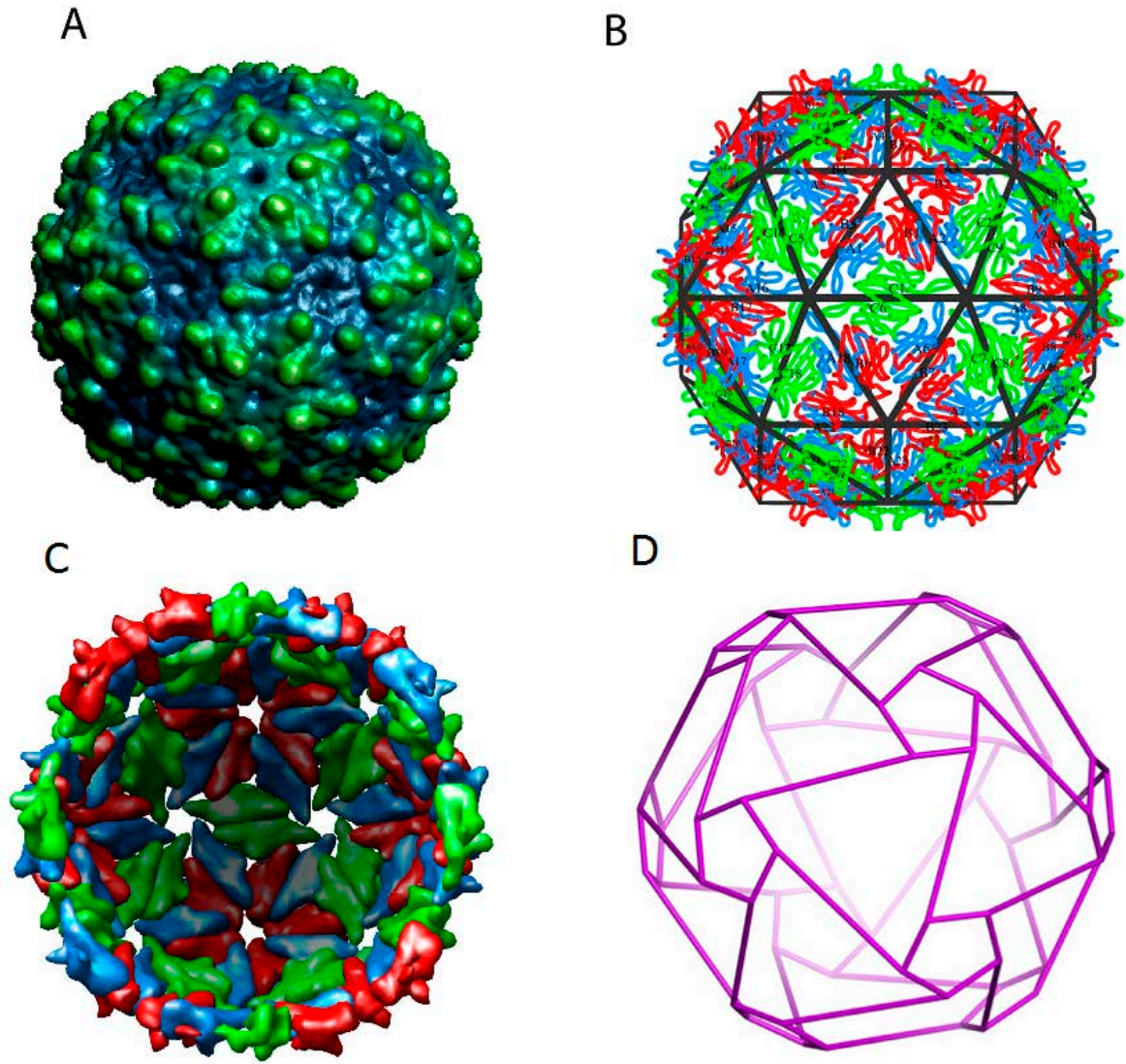

**Figure 4.** The genome packaging within MS2 capsid. (A) The surface overview of MS2 capsid. (B) The lattice representation of MS2 capsid. Each triangle contains one conformer A (blue), one conformer B (red), and one conformer C (green), which form dimers A/B and C/C. (C) The inside view of MS2 capsid. (D) The profiles of ssRNA density projected onto capsid surface obtained by cryo-EM (Toropova et al., 2008).

The genome of MS2 is of length $M = 3569$ (GenBank: NC_001417) and encodes coat protein, replicase, lysis and maturation protein, coat protein being the most highly expressed of four gene products. Using cryo-EM, Koning et al. (2003) and Toropova et al. (2008) determined RNA density profiles projected onto the capsid surface (see Fig. 4D taken from the paper by Toropova et al. (2008)) and elucidated some details of genome packaging in MS2. They proved that the genome in the wild-type virion is arranged in the form of two concentric shells with bimodal radial density distribution. These authors used averaging over icosahedral symmetry during cryo-EM data processing. The more recent studies (Dent et al., 2013; Koning et al., 2016; Dai et al., 2017) used asymmetric cryo-EM reconstruction of genome structure for phage MS2. The experiments by Koning et al. (2016) and Dai et al. (2017) resolved about 80% of RNA structure, including internal fraction of RNA previously attributed to bimodal radial density distribution, and proved the asymmetric compaction of RNA in the vicinity of maturation protein. We consider below how DFT and DDFT spectra can be applied to the interpretation of these experimental data and what can be learned from them.

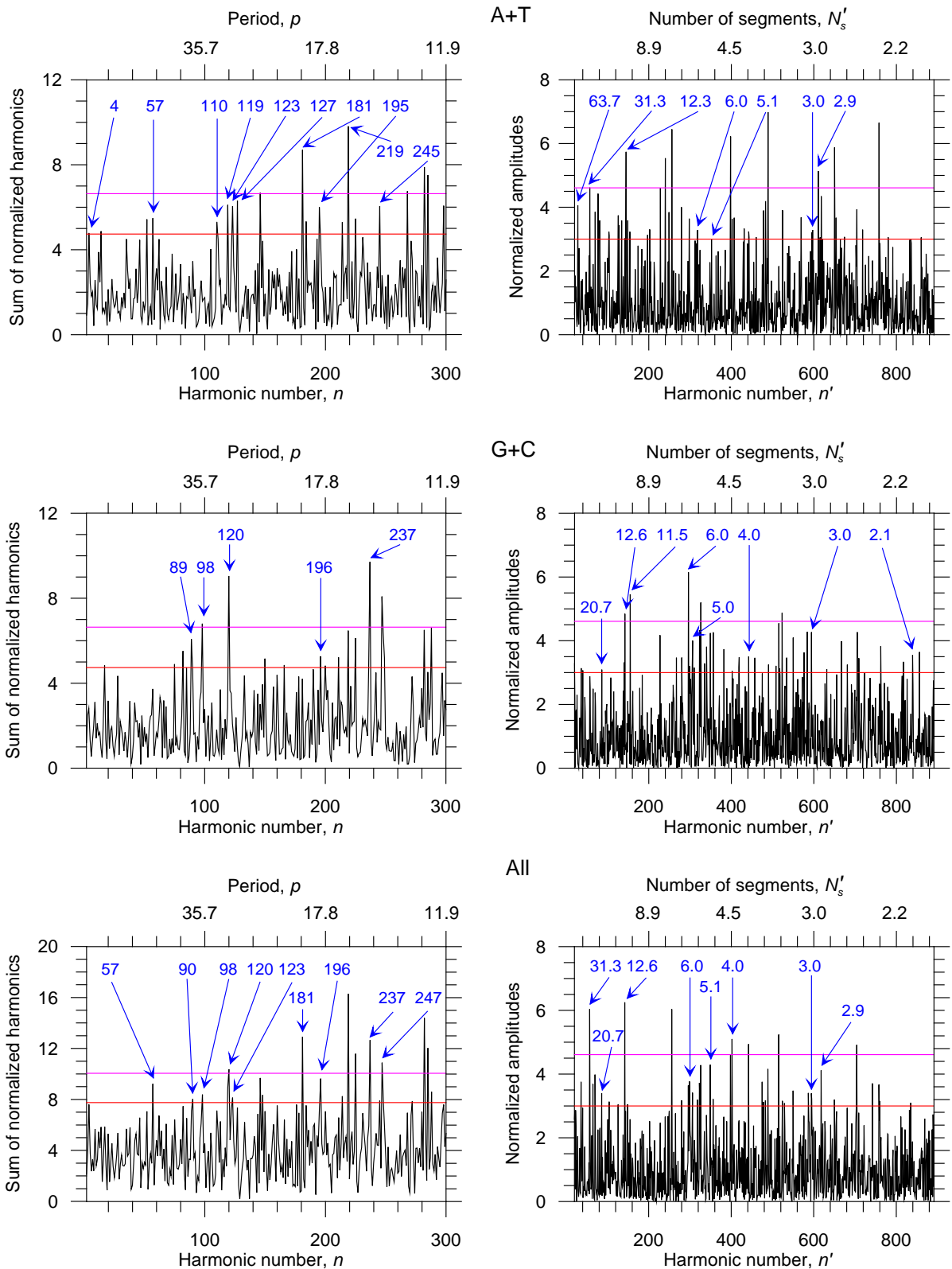

**Figure 5.** The DFT (left) and DDFT (right) spectra for the MS2 genomic sequence. The significant harmonics related to the elements of icosahedral symmetry are marked by arrows. The numbers of quasi-periodic segments corresponding to the marked harmonics are shown

explicitly in the panels. The red horizontal line corresponds to Pr = 0.05 for a particular harmonic, whereas the magenta horizontal line corresponds to Pr = 0.01.

The G+C DFT spectrum (Fig. 5, left; see also Supplementary file S1) shows clearly that the most pronounced large-scale segmentation corresponds to 120-mode (Pr = $1.18 \times 10^{-3}$). Again, as in the genomic sequence for STMV, the periodicities with high amplitudes were detected pairwise together with their doubled counterparts (harmonic with $n = 237$ in this case). The corresponding periodicities for A+T with $n = 110, 119, 123, 127$ (note also the counterpart doubled significant harmonics with $n = 219, 245$ for $n = 110, 123$) are less pronounced but clustered around $n = 120$. The probability to find 4 harmonics exceeding 5%-threshold within the range 110–130 is Pr = 0.02. Such correspondence is important for the general assessment of significance for 120 quasi-periodic segmentation. Despite the strong evolutionary divergence between STMV and MS2 and different (A+T)/(G+C) content ratio, the most pronounced 30-segmentation in STMV and 120-segmentation in MS2 appeared to be detected both in G+C DFT spectra. The related periods are also not far from each other (cf. $p = 29.7$ for MS2 and $p = 35.3$ for STMV). The elementary geometric consideration of RNA density profile on capsid icosahedral projection lattice (Koning et al., 2003; Toropova et al., 2008) proves that the sides of the triangles in Fig. 4D are approximately thrice longer the sides of the pentagons (for the perfect RNA profiles the correspondence would be exact). If the length of the pentagon side is defined as a unit length, the total length of projected RNA profile in Fig. 4D is equal to 120, that is in accordance with the distinct segmentation detected by DFT. In addition to 120-unit segmentation, we also found less pronounced periodic patterns with $n = 89$ for G+C and $n = 90$ for $S_4$ together with their doubled counterparts at $n = 181$ in A+T and $S_4$ DFT spectra. The 90- and 120-unit segmentation was also detected by the test with equidistant series in DDFT spectra (Supplementary file S2).

The correspondence between Fourier spectra and the asymmetric RNA tertiary structure is more complicated. Koning et al. (2016) observed 59 stem-loop units, of which 53 units ended near the capsid and the ends of 6 units were centrally located. The data by Dai et al. (2017) were similar, but the number of observed stem-loop units was a bit less. Despite the strong variations in unit lengths, the harmonics $n = 57$, $p = 62.6$ (A+T, $S_4$, DFT); $n = 52$, $p = 68.6$ (A+T, DFT) together with their approximate counterpart in DDFT spectra $N'_s \approx 63.7$ (A+T) provide the reasonable estimates of the total amount of stem-loop units. The symmetrical RNA profile in Fig. 4D seems to determine the general character of 120-segmentation in the MS2 genome, whereas the formation of the larger elements of RNA secondary structure can be performed by uniting such segments at the latter stages of maturation.

The genome for MS2 is 3.37 times longer the genome for STMV. If the periods of segments for MS2 would be the same as for 30-unit segmentation observed for STMV, the total number of segments in the genome of MS2 would be 101. The close segmentation with $n = 98$ together with its doubled counterpart at $n = 196$ was observed for G+C and $S_4$ DFT spectra (cf. also the harmonic with $n = 195$ for A+T DFT spectrum). Generally, the sets of quasi-periodic patterns with the periods in the range $p \approx 27$–$36$ were observed for both STMV and MS2 with the bias to the shorter periods for MS2.

The leading large-scale segmentation related to icosahedral symmetry detected by all DDFT spectra was $N'_s \approx 12$ (Fig. 5, right; note the clustering of significant harmonics around this value). This segmentation was also detected by the test with equidistant series in DDFT spectra (Supplementary file S2). The segmentation with $N'_s \approx 20$ turned out to be sufficiently less pronounced in comparison with $N'_s \approx 12$ and $N'_s \approx 30$. The segmentation with $N_s \approx N'_s \approx 60$ was also observed in DFT and DDFT spectra. The large-scale segmentation with $N'_s \approx 3$ dominates over segmentation with $N'_s \approx 5$ for A+T DDFT spectrum, whereas the significance of the both modes is comparable for G+C and $S_4$ DDFT spectra. Such a relationship between the segmentations $N'_s \approx 3$ and $N'_s \approx 5$ may reflect the competitive pathways in the capsid folding (Basnak et al., 2010).

### *3.3. Large-scale periodic patterns in ϕX174 genome*

The bacteriophage ϕX174 belonging to *Microviridae* family was chosen as an example of icosahedral viruses with ssDNA genome packaging (for a review see, e.g., Doore & Fane, 2016). The capsid of mature virus is composed of 60 copies each of the coat protein F, the spike protein G, the DNA-binding protein J, and 12 copies of the pilot protein H (PDB: 2BPA; McKenna et al., 1994; see also Figs. 6A–6C taken from VIPERdb). Its triangulation number is $T = 1$. The proteins H and J are associated with the inner capsid surface; the proteins H are beneath the centers of pentamers formed by the spike proteins G. There is one-to-one association between J and F proteins. The positions of J proteins are shown in Fig. 6C. The ϕX174 DNA-binding protein J is divided into three functional domains, 0, I, and II (Ilag et al., 1995; Bernal et al., 2004). The domains 0 and I are highly basic and positively charged that neutralize partially the negatively charged DNA. The domain II is very hydrophobic and contains no basic residues. The J proteins of the other *Microviridae* consist of only the domains I and II.

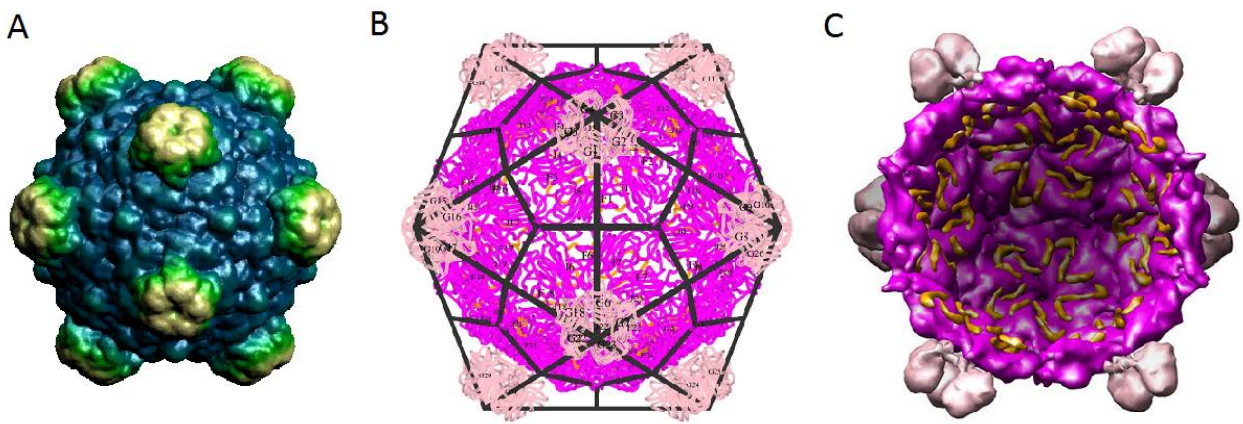

**Figure 6.** The structure of capsid for mature bacteriophage ϕX174. (A) The surface overview of ϕX174 capsid. (B) The dual icosahedron/dodecahedron representation of ϕX174 capsid. The spike proteins G are shown in light-brown, the coat proteins F in magenta, and the DNA-binding proteins J in yellow. (C) The inside view of ϕX174 capsid.

The genome of ϕX174 is of length $M = 5386$ and encodes 11 genes (GenBank: NC_001422). Although the ϕX174 genome is 1.5 times longer the MS2 genome, the inner radius of capsid for ϕX174 appears to be less than that for MS2 (9.6 nm versus 10.5 nm, respectively). Assuming the radius of ssDNA to be a half of that for dsDNA ($r_{dsDNA} \approx 1$ nm), the filling factor for the ϕX174 genome packaging can be estimated as 0.4. The icosahedrally ordered DNA is associated with domain I and part of domain II of the protein J (Bernal et al., 2004). The interactions with J protein tether ssDNA to the inner capsid surface. As ssDNA is concurrently synthesized and packaged along with the DNA-binding protein J, these interactions may perform also a scaffolding-like function during the procapsid-to-virion transition (Hafenstein & Fane, 2002).

The assembly and maturation of ϕX174 is a multi-stage process (McKenna et al., 1994; Door & Fane, 2016). The ssDNA bound with J proteins should be transported to procapsid, participates in the maturation and, finally, in the packaging itself. The transport stage implies one-to-one correspondence between ssDNA segmentation and J proteins. The relevant periodicities actually present in DFT and DDFT spectra (Fig. 7), $n = 61, 64$ (A+T, DFT); $n = 60, 68, 71$ (G+T, DFT); $n = 60, 71$ ($S_4$, DFT); and $N'_s = 61.2, 62.6, 58.5$ (G+C, DDFT). The 60-segmentation was also displayed in the test with equidistant series in DDFT spectra (Supplementary file S2).

The ssDNA-J protein associations and ssDNA secondary structure may be suggested to determine mainly the quasi-periodic segmentation related to the ϕX174 genome packaging. The interpretation of sequence analysis depends on the number of pinning contacts between ssDNA and J protein. The two pinning contacts would produce 120-segmentation. A ssDNA segment pinned by two contacts with J protein to the capsid surface should be bent in the middle that may be coordinated with the doubling of 120-segmentation. The pinning by each contact is associated with two ssDNA segments around contact. For this reason, there should be the geometric steric restrictions for ssDNA pinning to the inner capsid surface via 120 contacts, $240 S_{ssDNA} < S_{inner}$ (where $S_{ssDNA}$ is the cross-section area for ssDNA helix, $S_{ssDNA} \approx 0.78$ nm$^2$, and $S_{inner}$ is the inner capsid surface, $S_{inner} \approx 289.5$ nm$^2$ for $\phi$X174). As can be checked, this inequality is fulfilled. The pinning with regular quasi-periodic segmentation via 180 contacts would be hampered by the geometric steric restrictions. For the straight dsDNA stems the segmentation via 60 pinning contacts would provide the segments with half-lengths exceeding the inner radius of capsid. The choice between different alternatives needs additional experimental studies.

For A+T DFT spectrum the most salient large-scale mode refers to 120-segmentation (Fig. 7, left). There is the significant periodicity at $n = 118$ accompanied by the satellite high harmonics at $n = 109, 111, 113$, and 130. The probability to meet 4 harmonics exceeding 5%-threshold within interval $n = 110–130$ is Pr = 0.02. The counterpart significant harmonics at $n = 109$ and 130 were also detected for G+C and $S_4$ DFT spectra. The significant harmonics with the doubling of 120-segmentation were detected in both A+T and G+C DFT spectra, $n = 242$ (A+T) and $n = 239$ (G+C; the highest harmonic in related DFT spectral range). The test with equidistant harmonics in DDFT spectra revealed the broad spectrum of modes with 30-, 60-, 120-, and 240-segmentation, 120-segmentation being the most significant (Supplementary file S2).

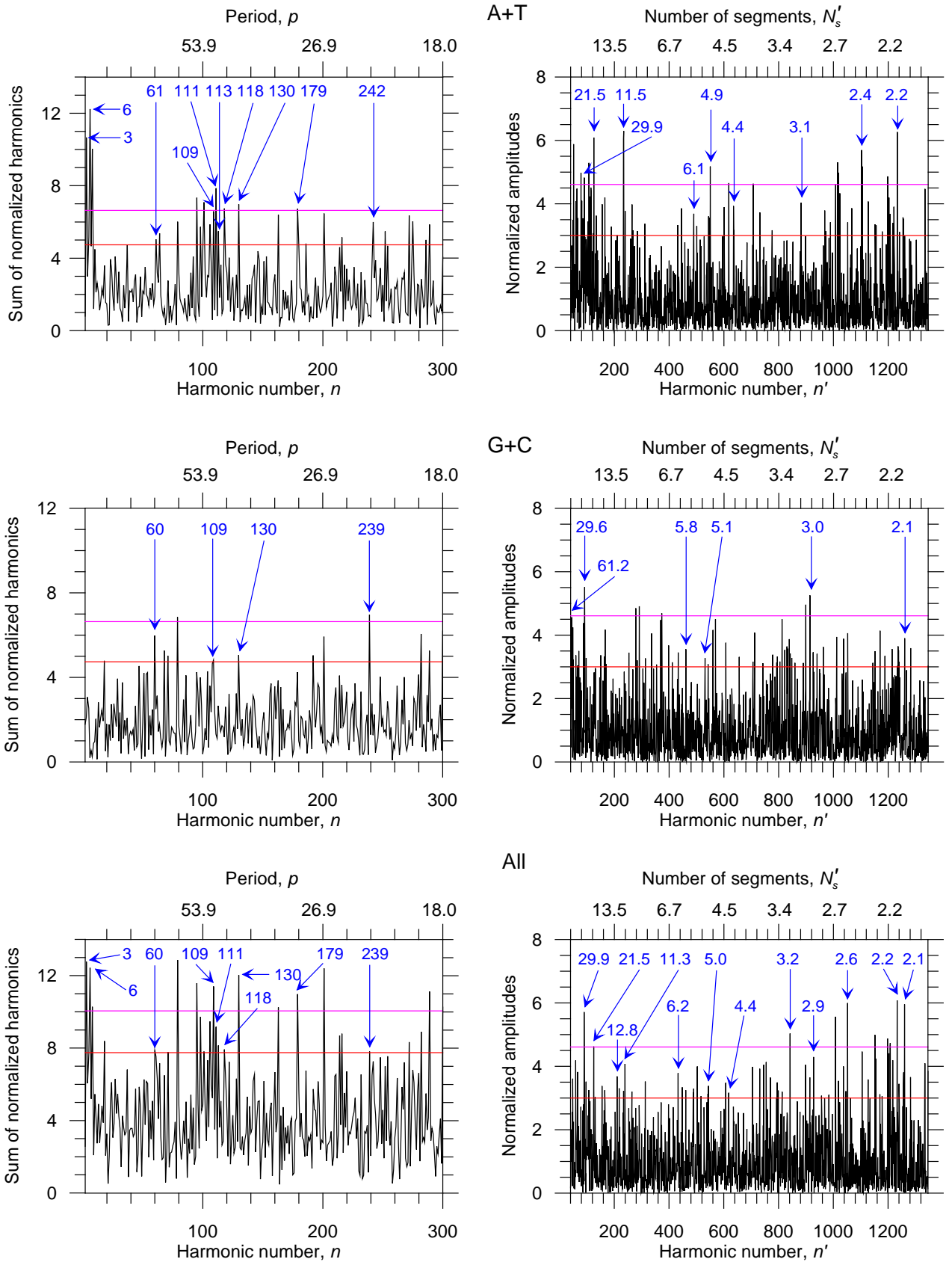

**Figure 7.** The DFT (left) and DDFT (right) spectra for the ϕX174 genomic sequence. The significant harmonics related to the elements of icosahedral symmetry are marked by arrows.

The numbers of quasi-periodic segments corresponding to the marked harmonics are shown explicitly in the panels. The red horizontal line corresponds to Pr = 0.05 for a particular harmonic, whereas the magenta horizontal line corresponds to Pr = 0.01.

Crystallographically, the form of the φX174 capsid is commonly attributed to the dodecahedron. Evidently, such attribution is not complete and the characterization of the φX174 capsid in terms of dual dodecahedron/icosahedron representation (Fig. 6B) is more convenient. The leading large-scale quasi-periodic segmentation detected in A+T DDFT spectrum (Fig. 7, right and Supplementary file S1) is associated with the faces/vertices in dual representation, $N'_s \approx 11.5$–$11.6$ (note also the commensurate high harmonic at $n' = 115$, $N'_s \approx 23.4$ related to the segmentation doubling for this mode). Such segmentation can be related to the hierarchical genome packaging via contacts with J proteins around 12 hydrophilic channels with the 5-fold symmetry formed by G and F proteins (Fig. 6C). Note also the possible relationship of this mode with the short A-T-rich repeating patterns corresponding to DFT harmonic $n = 576$, $p = 9.35$ (Chechetkin & Turygin, 1995). The corresponding harmonic is exactly commensurate with $N_s = 12$ (576/12 = 48). The other modes ordered by the amplitude ranking in A+T DDFT spectrum can be attributed to the vertices/faces ($N'_s \approx 21.5$) and to the edges ($N'_s \approx 29.9$). The leading mode in G+C DDFT spectrum corresponds to the edges ($N'_s \approx 29.6$; the highest harmonic in G+C DDFT spectral range, Pr = $4.01 \times 10^{-3}$). Note also the clustering of high harmonics around this mode.

The most large-scale segmentation modes with $N_s \approx N'_s \approx 3$ and $N_s \approx N'_s \approx 5$ were observed in both DFT and DDFT spectra. The latter segmentation was previously identified via technique with equidistant series in DFT spectra (Chechetkin & Turygin, 1995). These modes may be related to the hierarchical genome packaging at the initial stages of procapsid filling by ssDNA (presumably for $N_s \approx N'_s \approx 3$) and/or to the exit of ssDNA during host cell infection (presumably for $N_s \approx N'_s \approx 5$).

The results of Fourier analysis for the ϕX174 genome were compared with that for the Escherichia phage α3 genome (GenBank: NC_001330) of length $M = 6088$. The capsid assembly for the phage α3 (PDB: 1M06) is nearly the same as for ϕX174, except that J protein for α3 consist of only the domains I and II. The length of the genome for the phage α3, $M = 6087$, exceeds that for ϕX174 by the factor of 1.13. Therefore, the comparison of segmentations in the genomes of α3 and ϕX174 may shed light on how the genome length affects the packaging within capsids of the same geometric sizes. The summary of the results for α3 is briefly as follows. The periodicities $n = 67$ (A+T, $S_4$) and $n = 70$ (A+T, G+C, $S_4$) can be attributed to one-to-one correspondence with J proteins during transport of ssDNA-J protein complexes to procapsid (the bias from the expected value $n = 60$ may be related to both quasi-random variations in underlying segmentation and possible segregation of J proteins along ssDNA). The significant series multiple to $N'_s = 70$ were also obtained in the test with equidistant harmonics in DDFT spectra (Supplementary file S2). We found the shift of local maximum in A+T DFT spectrum toward $n = 240$. There are 5 harmonics exceeding the 5%-threshold within interval 230–250 (the probability for such event is Pr = 0.004) with the maximum at $n = 244$, $p = 24.9$ (Pr = $4.57 \times 10^{-4}$). The 240-segmentation for α3 is determined presumably by ssDNA secondary structure (perhaps, as subunits of the larger secondary structure units). Note the discrete character of transition from 120-segmentation in the genome of ϕX174 to 240-segmentation in the genome of α3 different from that would be expected according to the ratio of their genome lengths. Both 120- and 240-segmentation are compatible with icosahedral symmetry. The strict commensurability of the highly pronounced short-range periodicity at $n = 588$, $p = 10.35$ (A+T, DFT) for α3 with the number of dodecahedron faces (588/12 = 49) indicates its relevance to the genome packaging. This harmonic may be considered as a counterpart harmonic for ϕX174 at $n$

= 576, $p$ = 9.35 (A+T, DFT). DDFT spectra for α3 revealed the clustered significant harmonics associated with the edges, vertices, and faces of dual dodecahedron/icosahedron. The largest segmentation detected via DFT spectra corresponded to $n = N_s \approx 3$, whereas DDFT spectra indicated the leading segmentation $N'_s \approx 5$. The details of Fourier analysis for α3 can be found in Supplementary files S1 and S2.

### 3.4. Large-scale periodic patterns in SV40 genome

SV40 was chosen as a member of rapidly expanding *Polyomaviridae* family (for a review see, e.g., Atkin et al., 2009; Dalianis & Hirsch, 2013; Imperiale & Jiang, 2016). These small, non-enveloped, dsDNA viruses use mammals, birds, and fish as their natural hosts and may be the cause of different diseases including cancer. The 360 copies of the major virion proteins, VP1, in six conformations form 72 pentamers (or pentons) assembled in the capsid with $T = 7d$ lattice (PDB: 1SVA; Stehle et al., 1996; see also Fig. 8A taken from VIPERdb and Fig. 8B taken from Proteopedia; URL proteopedia.org; Prilusky et al., 2011). The minor virion proteins, VP2 or VP3, are located at the centers of pentons on the inner capsid surface. All virion proteins interact non-specifically with genomic dsDNA (Clever et al., 1993; Li et al., 2001; Roitman-Shemer et al., 2007; Tsukamoto et al., 2007; Oppenheim et al., 2008).

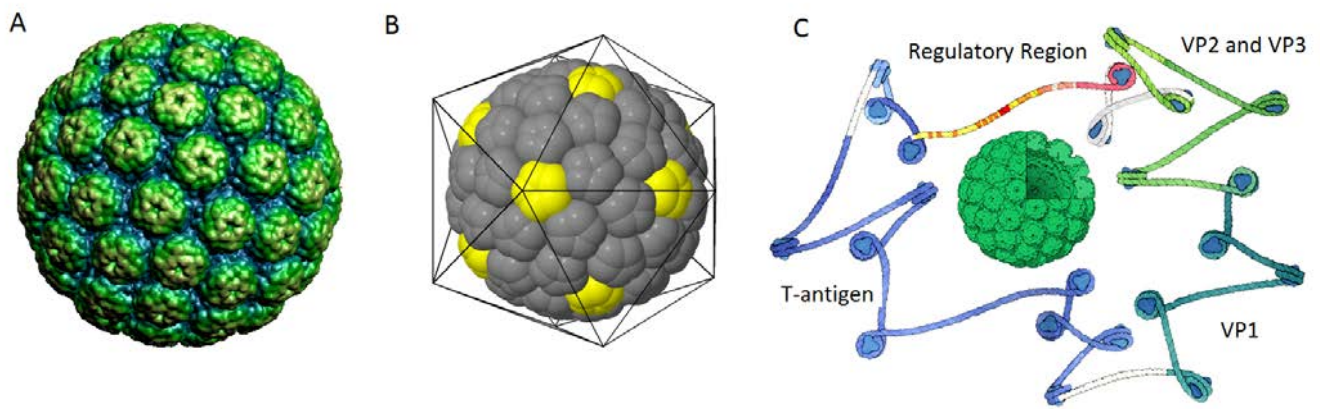

**Figure 8.** The structure of capsid for SV40. (A) The surface overview of SV40 capsid. (B) The schematic icosahedron representation of SV40 capsid. The pentons at the icosahedron vertices

are shown in yellow. (C) The minichromosome genome of SV40. The encoded genes are shown separately.

The SV40 genome of length $M= 5243$ (GenBank: NC_001669) encodes 8 genes (7 proteins and 1 RNA). The circular genome is organized into minichromosome with 20–26 nucleosomes (see Fig. 8C taken from PDB-101). The nucleosomes in the minichromosome packaged within the capsids of the mature virions are formed only of histones H2A, H2B, H3, and H4, whereas the histones H1 are absent, presumably due to compaction restrictions. The mean distance between the nucleosomes is 194–196 bp but may strongly vary depending on the mutations in VP1, environmental conditions, and virus life cycle (Ambrose et al., 1990). The mutations in VP1 may lead to the temperature-sensitive blocking of virion assembly (Blasquez et al., 1986; Ambrose et al., 1987; Behm et al., 1988). In particular, Ambrose et al. (1987) observed minichromosomes in mutant SV40 tsC219 with the period of 177 bp at 40°C with the same total number of nucleosomes as in the wild-type SV40. The shortening of period in mutant is due to the exposed regulatory region free from nucleosomes (Ambrose et al., 1990; and references therein). Coca-Prados et al. (1982) reported the average periods of 211 bp for virion assembly intermediates.

The inner radius of capsid for SV40 is 17.9 nm. The free dsDNA of length $M= 5243$ without nucleosomes would occupy only of 0.23 capsid volume. Estimating the volume of nucleosome without H1 as about 440 nm$^3$ and the filling factor as 0.6, the number of nucleosomes within capsid may amount to 20 that is close to the experimentally observed number. Keller et al. (2009) studied SV40 chromatin structure via cryo-electron tomography, however, the organization of the viral chromatin has not been determined. Saper et al. (2013) using small-angle X-ray scattering proved that minichromosomal density increases toward the center of the capsid. Recently, Hurdiss et al. (2016) solved the structure of human BK polyomavirus at 7.6 Å providing insights into the location of minor capsid proteins, genome recognition, and organization of the viral minichromosome. They observed two-shell electron density adjacent to the inner capsid layer and the maximum of density at the center. The nucleosomes have not been resolved in this study as well. We will discuss below how all these features can be related to the periodic patterns detected in SV40 genome.

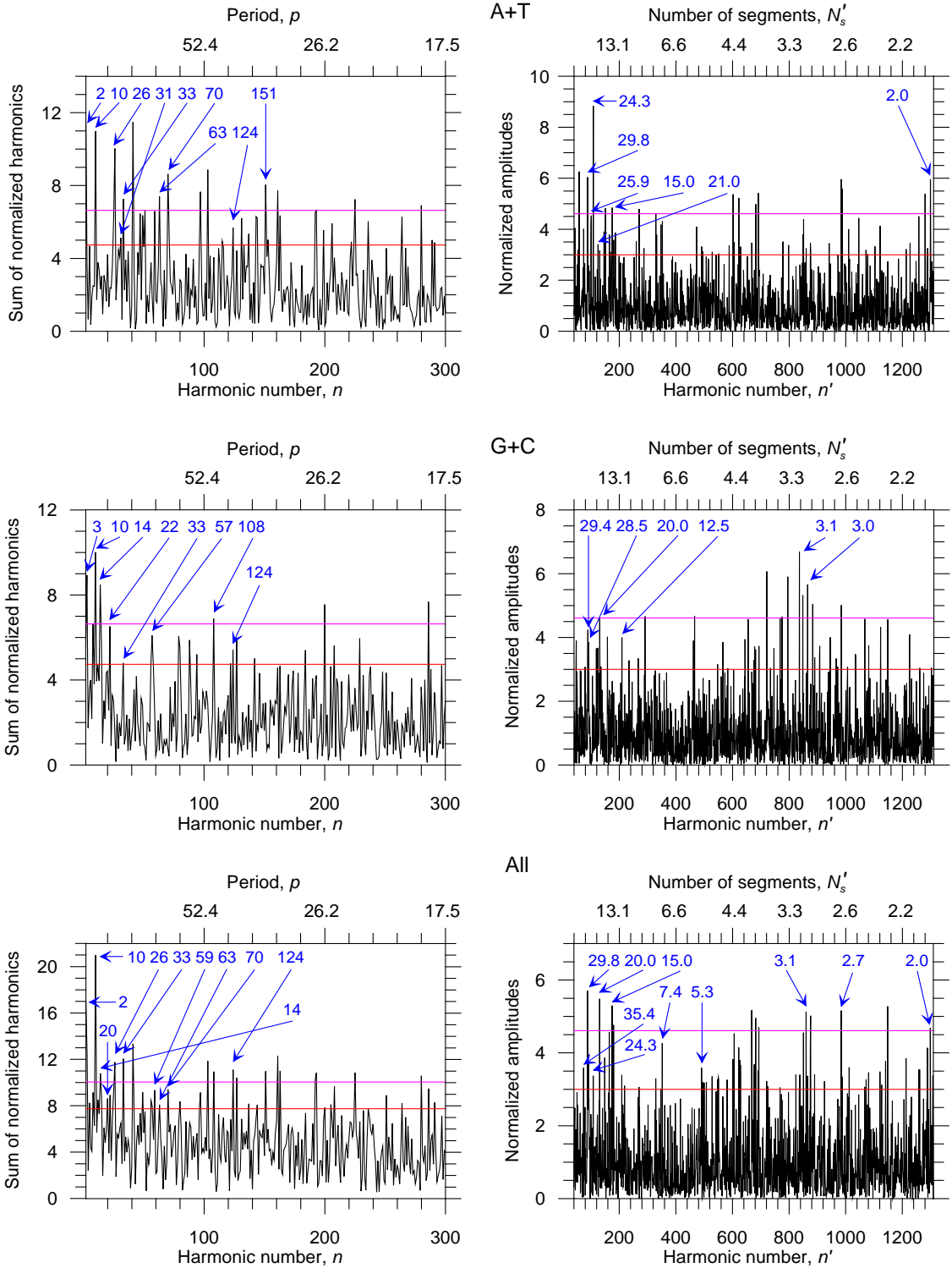

**Figure 9.** The DFT (left) and DDFT (right) spectra for the SV40 genomic sequence. The significant harmonics related to the positioning of nucleosomes and to the elements of icosahedral symmetry are marked by arrows. The numbers of quasi-periodic segments corresponding to the marked harmonics are shown explicitly in the panels. The red horizontal line corresponds to Pr = 0.05 for a particular harmonic, whereas the magenta horizontal line corresponds to Pr = 0.01.

The corresponding DFT and DDFT spectra are shown in Fig. 9 (see also Supplementary file S1). As described above, there is a broad spectrum of distances between nucleosomes on minichromosome. All related periods are seen in both DFT and DDFT spectra. The set of significant nucleosome periods includes the harmonics (all ranked by the amplitude height): DFT, A+T, $n = 26$, $p = 201.7$; $n = 33$, $p = 158.9$; $n = 25$, $p = 209.7$; $n = 31$, $p = 169.1$; G+C, $n = 33$, $p = 158.9$; $S_4$, $n = 33$, $p = 158.9$; $n = 26$, $p = 201.7$; $n = 25$, $p = 209.7$; DDFT, A+T, $N'_s = 24.3$, $p' = 216.1$ (the highest harmonic in A+T DDFT spectral range, Pr = $1.45 \times 10^{-4}$); $N'_s = 29.8$, $p' = 176.1$ (the third by ranking in A+T DDFT spectrum); $N'_s = 25.9$, $p' = 202.1$; $N'_s = 26.2$, $p' = 200.1$; G+C, $N'_s = 29.4$, $p' = 178.1$; $N'_s = 28.5$, $p' = 184.1$; $S_4$, $N'_s = 29.8$, $p' = 176.1$ (the highest harmonic in $S_4$ DDFT spectrum, Pr = $3.33 \times 10^{-3}$); $N'_s = 28.8$, $p' = 182.1$; $N'_s = 29.4$, $p' = 178.1$; $N'_s = 24.3$, $p' = 216.1$. For the purposes of our paper, it is interesting to note that the period $p = 177$ revealed previously in the temperature-sensitive SV40 mutants (Ambrose et al., 1987) appeared to be closely associated with 30-segmentation, moreover, this mode turned out to be the third by ranking in A+T DDFT spectrum and the highest in $S_4$ DDFT spectrum (note also the significant harmonics with $n = 31, 33$ in DFT spectra). In both DFT and DDFT spectra there are clear trends indicating the cascade of period-doubling for 30-segmentation (i.e., the consecutive segmentation on 15 and 7.5–8 segments), which may be treated as a trend to the formation of di- and four-nucleosome units in the set of nucleosomes with the periods $p \approx 176$–177. These features indicate the important role of DNA association with the capsid edges during the genome packaging. It is well-known that the nucleosome positioning is closely related to the particular di- and/or oligonucleotides phased with dsDNA helix pitch (Clark, 2010; Trifonov, 2011). We observed the related high harmonics with periods $p \approx 10.0$–$10.4$ in both A+T and G+C DFT spectra.

Among the other significant modes potentially associated with icosahedral symmetry, we found the harmonics with $N_s \approx N_s' \approx 20$, 12, and 60. We also checked multiple 72-segmentation that may be relevant in the case of multiple DNA-capsid protein contacts. The significant modes with $N_s \approx N_s' \approx 72$ and 144 were detected in DFT spectra and in the test with equidistant series in DDFT spectra (Supplementary files S1 and S2). The significant most large-scale segmentation detected by both DFT and DDFT spectra corresponded to $N_s \approx N_s' \approx 2$ and 3. All these large-scale quasi-periodic patterns in SV40 genome indicate cooperative DNA-protein interactions in the genome packaging and/or virion assembly.

### 3.5. Large-scale periodic patterns in HK97 genome

The dsDNA bacteriophages belong to the most abundant species in the nature (Calendar & Abedon, 2006; Rossmann & Rao, 2012; Mateu, 2013). Despite essential variations in their genome lengths, ~20–170 kb, the general architecture of the genome packaging resolved by cryo-EM for many tailed dsDNA bacteriophages appears to be conservative (for a review see, e.g., Johnson & Chiu, 2007). Unlike the viruses considered above, for which the virions are self-assembled, the dsDNA packaging into capsids of tailed bacteriophages needs special molecular machinery (Feiss & Rao, 2012). Black & Thomas (2012) described eight distinct models of the genome packaging within capsids of tailed bacteriophages proposed by different authors and this list can be prolonged even further. In this section we present some results concerning the large-scale periodic patterns in the genomic sequence for the tailed dsDNA bacteriophage HK97. A part of the features observed is expected to be generic, whereas the other part may vary for different species.

The protein gp5, together with the protease gp4, guides the assembly and maturation of HK97 capsid (PDB: 1OHG; Wikoff et al., 2000; Helgstrand et al., 2003; Suhanovsky & Teschke, 2015; see also Figs. 10A–10C taken from VIPERdb). When expressed alone, 420 copies of protein gp5 assemble into a portal-deficient capsid with $T = 7I$ lattice via several stages of maturation. The proteins gp5 form 60 hexamers and 12 pentamers. During the assembly of tailed dsDNA bacteriophages, one of the pentamers is replaced by a 12-mer of the portal protein gp3, that constitutes the gateway for DNA packaging.

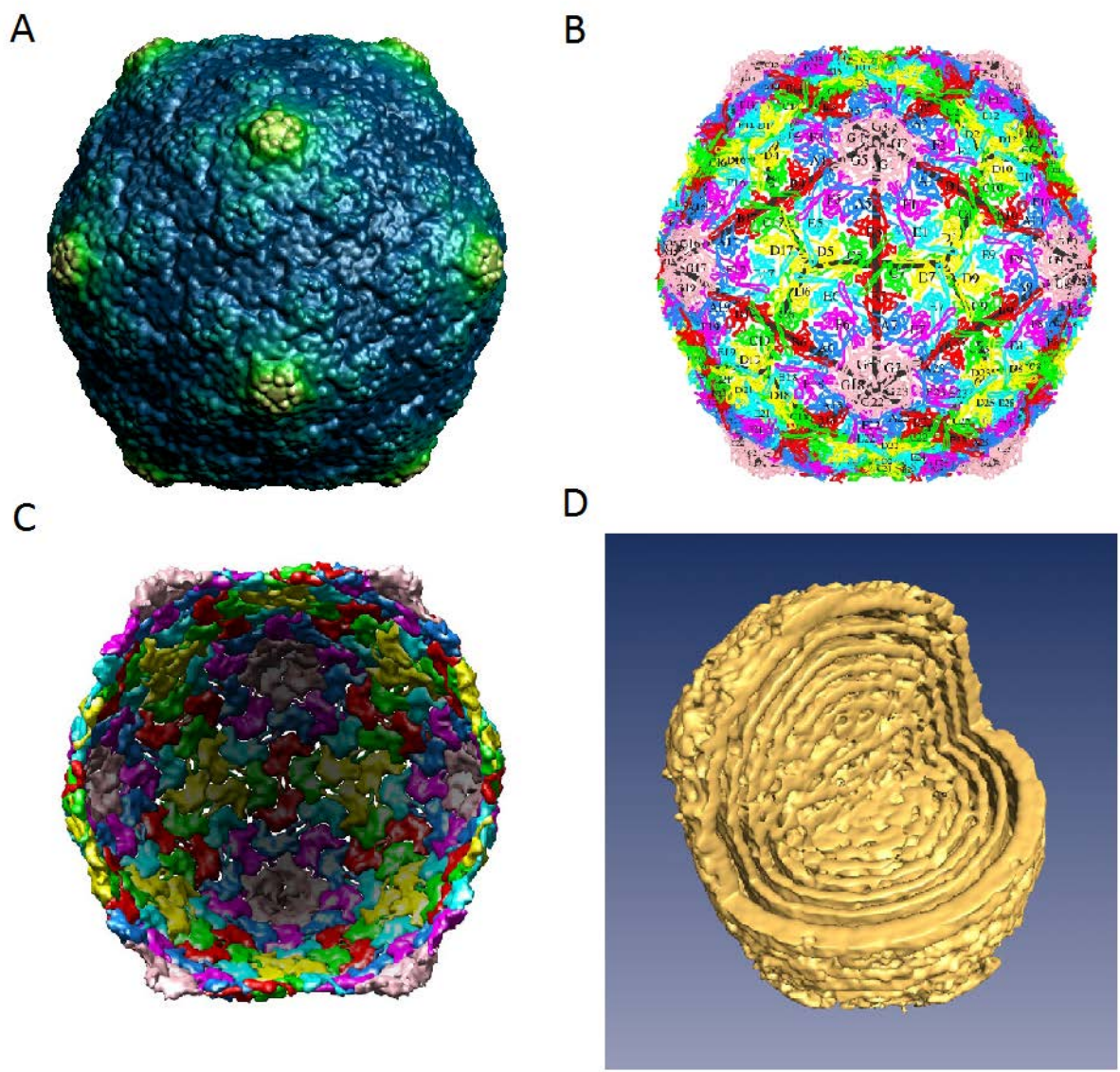

**Figure 10.** The structure of capsid and genome packaging for dsDNA bacteriophage HK97. (A) The surface overview of HK97 capsid. (B) The dual icosahedron/dodecahedron representation of HK97 capsid. The 7 different proteins gp5 in asymmetric unit cell are shown in different colors.

(C) The inside view of HK97 capsid. (D) The cryo-EM micrograph illustrating genome packaging within ϕ29 capsid (Comolli et al., 2008). The genome of HK97 is packaged similarly (Duda et al., 2009).

The cryo-EM micrographs revealed that dsDNA within icosahedral capsids of many tailed bacteriophages is ordered into a set of concentric layers with spacing ~2.5 nm (Johnson & Chiu, 2007; Duda et al., 2009; see also Fig. 10D taken from the paper by Comolli et al., 2008). The bending rigidity of dsDNA is characterized by persistence length, $L_{pers} \approx 150$ bp (or 51 nm). The other important characteristic of dsDNA is related to the Kuhn length, $L_K$. The segments with lengths exceeding $L_K$ may be subject to essential bending deformations and at this distances dsDNA may be approximately considered as a freely articulated chain. The persistence and Kuhn lengths are related as $L_K \approx 2L_{pers}$ (Grosberg & Khokhlov, 1994). These estimates show that the optimal characteristic length $L_{pers} < L_{ch} < L_K$ should exist, which can be associated with a characteristic element of segmentation in genomic dsDNA. The estimates in this topic may also be addressed to the virus SV40 considered in the previous section, in which there are the additional steric restrictions related to dsDNA winding on the nucleosomes. We will show below that the segmentation with the characteristic length(s) $\sim L_{ch}$ plays the important role in the HK97 genome that is presumably the generic feature for the other dsDNA bacteriophages.

The linear genome of HK97 is of length $M = 39732$ (GenBank: NC_002167) and encodes 62 genes and 1 pseudogene. The corresponding DFT and DDFT spectra for this genomic sequence are presented in Fig. 11 (see also Supplementary file S1). The relevant DFT spectral range reveals a trend at the low spectral numbers $n$ reflecting the mosaic patchiness of the HK97 genome (related to multiple gene coding, large-scale variations in A/T and G/C distribution, etc.). As these large-scale effects are not necessary quasi-periodic, the additional application of DDFT is crucial for the assessment of true periodicity in such cases (Chechetkin & Lobzin, 2017).

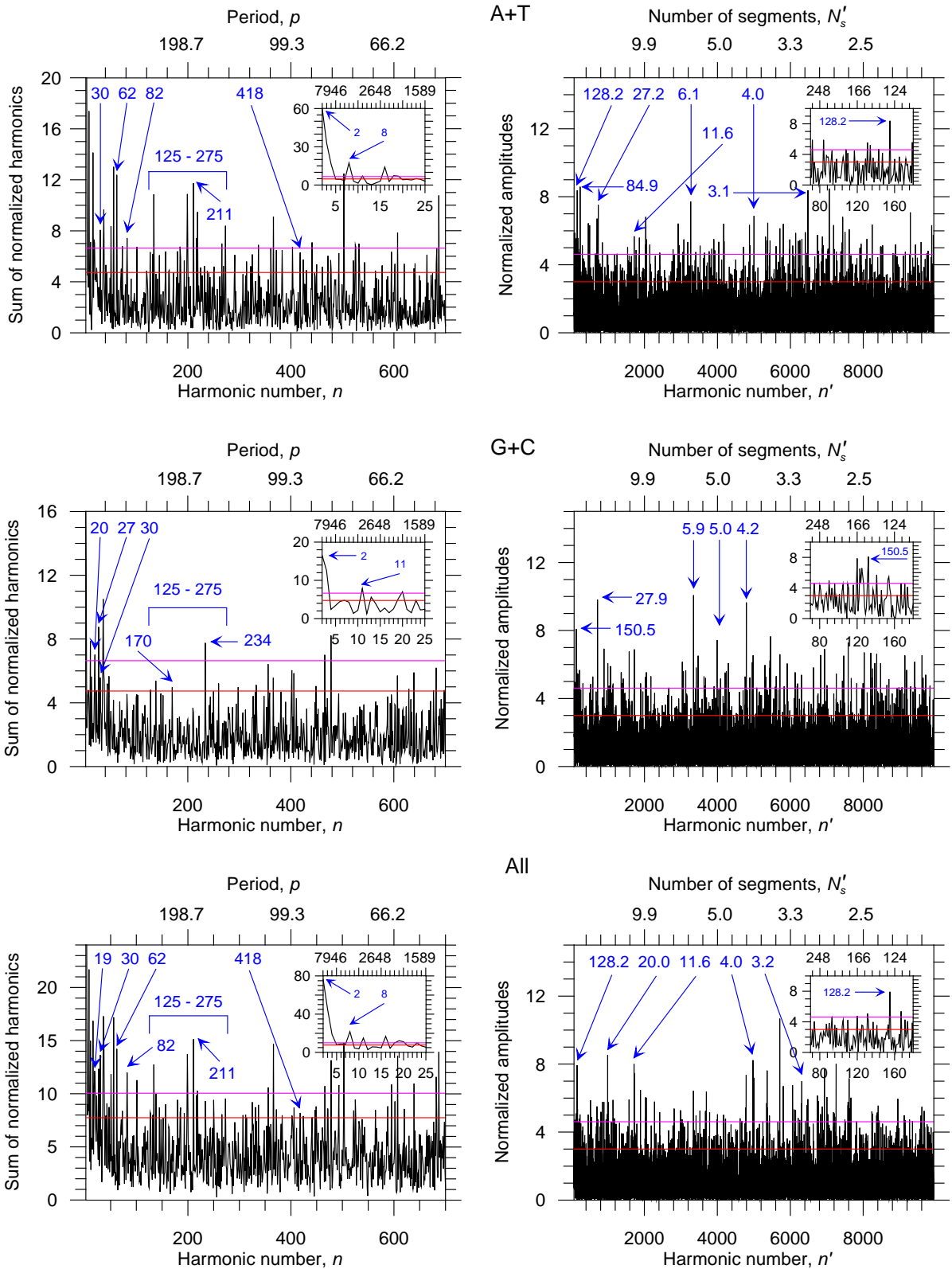

**Figure 11.** The DFT (left) and DDFT (right) spectra for the HK97 genomic sequence. The significant harmonics related to the elements of icosahedral symmetry are marked by arrows. The numbers of quasi-periodic segments corresponding to the marked harmonics are shown explicitly in the panels. The spectral ranges corresponding to the segments with periods between the persistence and Kuhn lengths are marked by overbar on DFT spectra (left) and shown separately in the inserts on DDFT spectra (right). The inserts on DFT spectra illustrate a trend at the low spectral numbers reflecting the mosaic patchiness of the HK97 genome. The red horizontal line corresponds to $Pr = 0.05$ for a particular harmonic, whereas the magenta horizontal line corresponds to $Pr = 0.01$.

Beyond the trend range, there was a significant enrichment of range $n = 125–275$, $p = 145–318$ by the high harmonics in A+T DFT spectrum. The number of harmonics exceeding the 5%-significance threshold in this range was 24 (at the expected number in random spectra $7.55\pm2.75$; the probability of such event is $Pr = 9.98\times10^{-7}$). Such enrichment was absent in the counterpart range for G+C DFT spectrum. The similar enrichment of this spectral DFT range for $S_4$ can be mainly attributed to A+T contribution. The highest harmonic in the range $n = 125–275$ for A+T and $S_4$ DFT spectra was at $n = 211$, $p = 188.3$ (remind that $n = N_s$ for DFT). The enrichment of range $n' = 73–159$, $p' = 146–318$ by the significant harmonics was observed in all A+T, G+C, and $S_4$ DDFT spectra. The total numbers of harmonics exceeded the threshold corresponding to $Pr = 0.05$ were 20, 31, and 18 (at the expected number in random spectra $4.35\pm2.09$). The highest harmonics in this DDFT range were at $n' = 155$, $N'_s = 128.2$, $p' = 310.0$ for A+T and $S_4$ ($Pr = 2.28\times10^{-4}$ and $3.59\times10^{-4}$, respectively; note the correspondence to the significant counterpart harmonics with $n = 127$ in DFT spectra); and $n' = 132$, $N'_s = 150.5$, $p' = 264.0$ for G+C ($Pr = 3.05\times10^{-4}$). The test with equidistant series in DDFT spectra indicates the significance of segmentation with $N'_s = 150$ and 210 (Supplementary file S2).

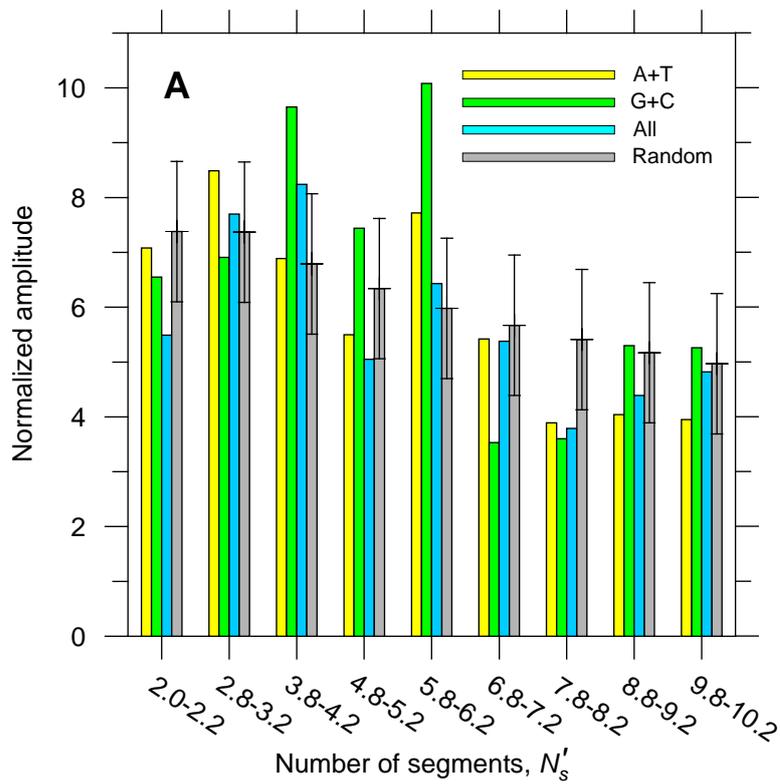

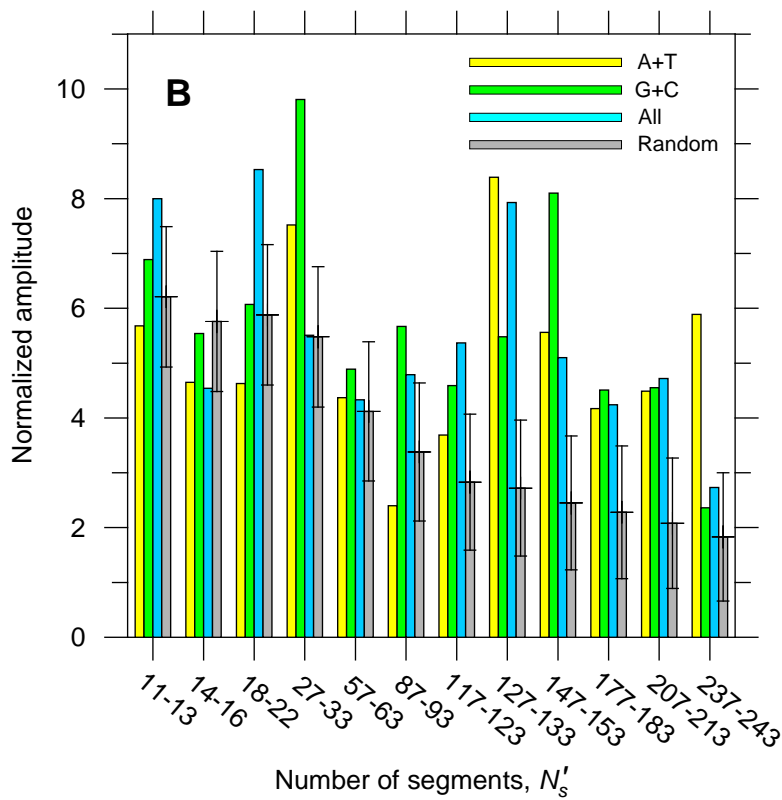

**Figure 12.** The highest harmonics in the DDFT spectral ranges corresponding to the longer (A) and shorter (B) segmentation modes in comparison with their counterparts in the random spectra assessed by extreme value statistics.

The assessment of significance in particular DDFT spectral ranges containing many harmonics (as in the case of HK97) needs the application of extreme value statistics (Lobzin & Chechetkin, 2000). In the random spectra, the harmonics with the different spectral numbers $n'$ are statistically independent and are related reciprocally with the number of segments $N'_s$ (Eq. (15)). Therefore, the harmonics corresponding to segmentation with small $N'_s$ will have the larger statistical weight and the larger probability to detect the high harmonics in this range. We assessed the segmentation in the different ranges taking into account the effects of extreme value statistics. The results are summarized in Figs. 12A–12B. The threshold $<f_{max}>_{random} + 1.96\sigma(f_{max})$ (where $<f_{max}>_{random}$ is the mean maximum amplitude in a range of random spectrum with the same number of harmonics as in the range under study and $\sigma(f_{max})$ is the corresponding standard deviation) was exceeded by the following harmonics: $N'_s \approx 4$ (G+C); $N'_s \approx 6$ (G+C); $N'_s \approx 20$ ($S_4$); $N'_s \approx 30$ (G+C); $N'_s \approx 120$ ($S_4$); $N'_s \approx 128-129$ (A+T, G+C, $S_4$); $N'_s \approx 150$ (A+T, G+C, $S_4$); $N'_s \approx 210$ (A+T, G+C, $S_4$); $N'_s \approx 240$ (A+T). The segmentation modes $N'_s \approx 128-129$ and $N'_s \approx 150$ proved to be the most significant by this criterion as well (note also the significance of the mode $N'_s \approx 210$).

A part of significant periodicities detected by DFT and DDFT may be attributed to the DNA packaging within layers. For a spherical-like packaging the length of repeating patterns can be estimated as $l_L \approx 2\pi R_L$, where $R_L$ is the radius of the layer. For the radius of the outer layer about 24.6 nm, this yields $l_L \approx 154.6$ nm or 454.6 bp. In the case of fixed spacing between the layers, $\Delta R_L \approx 2.5$ nm, the length of repeats in the adjacent layers would be changed on $\Delta l_L \approx 2\pi\Delta R_L$ or ~45.9 bp. The series of periods associated with concentric DNA layers can be extended up to $l_L \approx 150$ bp. Such series were detected in both DFT and DDFT spectra. The highest harmonic in A+T DDFT spectrum was at $n' = 234$, $N'_s = 84.9$, $p' = 468.0$ (Pr = $1.85\times10^{-4}$) and was significant in G+C and $S_4$ DDFT spectra (note also the approximately related harmonics at $n = 82$ in A+T and $S_4$ DFT spectra). This harmonic was accompanied by the exactly doubling segmentation $n' = 117$, $N'_s = 169.8$, $p' = 234.0$ in A+T DDFT spectrum and has counterparts $n' = 116$, $N'_s = 171.3$, $p' = 232.0$ and $n = 170$, $p = 233.7$ in G+C DDFT and DFT spectra. These features may be related to the DNA packaging within outer layers.

**4. Discussion**

Some of the detected quasi-periodic patterns are distinctly associated with RNA/DNA-capsomer interactions and depend on the assembly and subsequent packaging mechanisms. Based on the experimental data for STNV and MS2, Stockley et al. (2016) developed two-stage model for the assembly of ssRNA viruses. At the first, more rapid, stage RNA binds to coat proteins to facilitate capsid assembly, whereas at the second, slower, stage RNA is compactly packaged within capsid. The specific cooperative RNA-coat protein interactions play important role at the both stages. As the motifs responsible for specific recognition differ generally at the two stages, we reassign the terminology by Stockley et al. (2016) and will call the specific RNA motifs at the first stage as assembly signals (AS), while the motifs at the second stage will be called as packaging signals (PS). The model by Stockley et al. (2016) implies generally RNA refolding at the two stages.

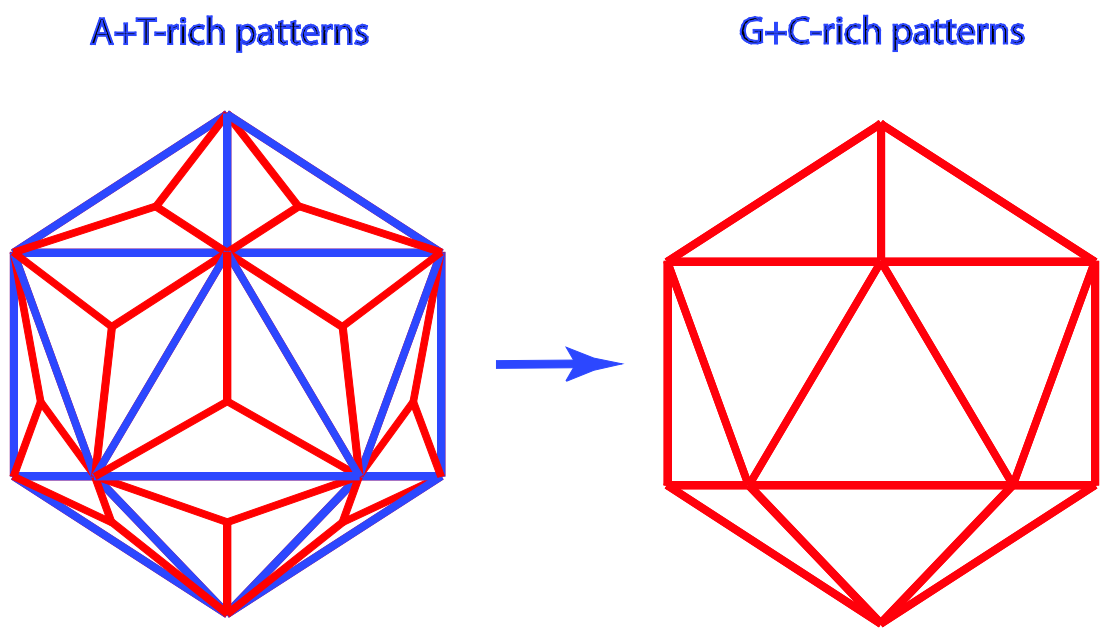

**Figure 13.** The scheme of two-stage packaging mechanism for STNV suggested by Stockley et al. (2016). The mode on the left corresponds to the nearly completed first stage of the virion assembly. The resulting packaging mode on the right at the completion of the second stage is based on the experimental data by Larson et al. (2014) for STMV (see also spectra in Fig. 2). The RNA profile projected onto the capsid surface is shown in red.

The transition from the first to second stage for STMV and STNV can be presented schematically as shown in Fig. 13. Patel et al. (2017) suggested for the first assembly stage the motifs $AN_2A$. One-to-one correspondence between these motifs and coat proteins implies 60-segmentation with A+T-rich patterns. The data by Larson et al. (2014) unambiguously assign the strikingly pronounced harmonic at $n = 30$ (G+C, DFT) in Fig. 2 to 30 stem-loop units along icosahedron edges. This means that 30-segmentation at the second packaging stage is mainly associated with G+C-rich periodic patterns. The significant harmonics in the vicinity of $n = 60$ in A+T DFT spectrum may be related to AS suggested by Patel et al. (2017). The two-stage transition shown in Fig. 13 agrees with Fourier spectra for STMV and STNV-C, while the mode $n = 30$ disappears in G+C DFT spectrum for STNV-2 (see Fig. 3) and is strongly distorted and suppressed in G+C DFT spectrum for STNV-1 (see Supplementary file S1). The change in Fourier spectra correlates clearly with the genome lengthening. This raises a question: could AS serve simultaneously as "backdoor" PS at the genome lengthening? Indeed, if the length of icosahedron edge is fixed, the rearrangement in resulting genome packaging from the mode shown on the right to that shown on the left in Fig. 13 permits the packaging of genome approximately $2/\sqrt{3} \approx 1.155$ longer. The relative lengths for the STNV genomes are $M/M_{STMV} =$ 1.154 (STNV-C), 1.171 (STNV-1), and 1.177 (STNV-2) and are not far from $2/\sqrt{3} \approx 1.155$. Therefore, the change in Fourier spectra may indicate the transition to 60-segment packaging mode (or to the frozen first stage) for STNV-1 and STNV-2. The experimental observation of 60-segment packaging mode for STNV-2 (Fig. 10, left) instead of 30-segment packaging mode for STMV (Fig. 10, right) via cryo-EM or X-ray techniques would be of basic interest for understanding of the assembly abilities of ssRNA viruses. Alternatively, the resulting 30-segmentation packaging mode may also be retained for STNV-2 but with A+T-rich motifs (see Fig. 3, left). In the latter case it should be clarified how the resulting 30-segmentation packaging mode with A+T-rich motifs can be reconciled with the length of STNV-2 genome. In a fuzzy form the similar rearrangement of packaging might tackle the insertions within UTR part of the STMV genome.

The harmonics in Fourier spectra associated with AS at the first assembly stage for MS2 should correspond to the number of dimers and may be attributed to $n = 89$, $p = 40.1$ (G+C, DFT); $n = 90$, $p = 39.7$ ($S_4$, DFT) (note also the doubling of this segmentation, $n = 181$, $p = 19.7$ (A+T, $S_4$, DFT)). The 90-segmentation was also identified via the test with equidistant harmonics in DDFT spectra (Supplementary file S2). The most pronounced 120-segmentation should evidently be attributed to the second packaging stage. Such correspondence was obtained for RNA profile determined by the data averaged over icosahedral symmetry (Koning et al., 2003; Toropova et al., 2008). There is clear similarity between patterns for 120-segmentation in the genome of MS2 and 30-segmentation for STMV. The reconstruction of motifs for harmonics $n = 89$ and $n = 120$ yields $N_4SN_6GN_7cymyN_7(\text{non-c})N_4rN_3r$ (italic letters correspond to $0.05 < \text{Pr} < 0.06$) and $cN_4wN_3(\text{non-}a)N_{11}tcN(\text{non-}g)N_2mN_2$ (where by the bias in respective frequencies, w ≈ a, m ≈ c, and non-g ≈ c). Notably, the motif for the harmonic $n = 57$, $NYN_6aakNnon-cN_{12}rN_5aTNAN_6cmN_3non-cN_5GN_4mNGgN_2$, contains AS $AN_2A$ typical of STNV. Although asymmetric cryo-EM reconstruction provides distinctly different picture (Koning et al., 2016; Dai et al., 2017), Dent et al. (2013) proved that asymmetric profile still retains the approximate features inherent to icosahedral symmetry. Note that the total amount of stem-loop units observed by asymmetric cryo-EM reconstruction corresponds crudely to a half of 120-segmentation. Though the inclusion of the maturation protein into the capsid breaks the complete icosahedral symmetry, the symmetry restriction should be important at least at the first assembly stage. The pronounced 120-segmentation indicates that icosahedral symmetry works somehow even in the case of the broken symmetry at the second packaging stage. For both ssRNA viruses, STMV and MS2, we found also the large-scale segmentation reflecting the hierarchical genome organization and possibly the packaging via 3-fold or 5-fold intermediates. These features are absent in non-viral RNA, which may also support assembly but inefficiently and with aberrant structures.

The maturation and packaging of ssDNA viruses ϕX174 and α3 is multi-stage, the transport and packaging stages being separated. The transport of ssDNA bound to J proteins to procapsid implies one-to-one correspondence between related segmentation and the number of J proteins (see, e.g., McKenna et al., 1994). The segmentation related to the packaging itself depends on ssDNA association with J proteins and ssDNA secondary structure. The corresponding segmentation compatible with icosahedral symmetry was detected in both ϕX174 and α3 genomes. The shorter segmentation in the α3 genome in comparison with that for ϕX174 may facilitate the packaging of the longer α3 genome within capsid.

For SV40 we found that quasi-periodic segmentation in genomic sequences reflects the minichromosome organization and nucleosome positioning. The dynamic rearrangement of nucleosomes during virus life cycle includes the segmentation $N_s \approx N'_s \approx 30$ concordant with the icosahedral symmetry and the genome packaging within capsid. The experiments yield commonly the lower numbers for nucleosomes, ~20–26. The packaging with the such numbers of nucleosomes cannot be in a complete correspondence with the icosahedral symmetry and a part of randomization in nucleosome positioning within capsid observed experimentally (Keller et al., 2009; Saper et al., 2013; Hurdiss et al., 2016) can be attributed to such incomplete correspondence. By analogy with the cryo-EM asymmetric reconstruction for MS2, the results by Saper et al. (2013) and Hurdiss et al. (2016) indicate the presence of internal fraction in DNA packaging. Such packaging mode can in part be related to the repulsion of positively charged histones from capsid. Our results indicate also the cooperative dsDNA-penton interactions.

Perhaps, the two-stage model by Stockley et al. (2016) may be extended to this class of viruses as well. In this case dsDNA-penton interactions may be related to the first assembly stage. The reconstruction of motif corresponding to the leading strand for the harmonic $n = 70$ yields $N_8cN_2tN_{11}aNGsN_2(\text{non-g})N_4wtN_2mN_6wrN_3tTN_{10}sYN_7kN_2t$ ($Y \approx c$, $m \approx a$).

The segmentation related to the bending rigidity of dsDNA appeared to be the most salient for HK97 genome. The quasi-periodic patterns in the range $p = p' = 146–318$ with $N_s \approx N'_s \approx 150$ and $N_s \approx N'_s \approx 210$ may have also the relevance to the icosahedral symmetry. According to cryo-EM micrographs the form of each DNA layer reproduces up to the similarity transform the form of capsid indicating the coordination between icosahedral symmetry and the form for each layer. Therefore, the correspondence with icosahedral symmetry remains global despite the multi-layer mode of the genome packaging for dsDNA bacteriophages.

Although the spool-like genome packaging with transverse axial symmetry is often taken for granted for dsDNA bacteriophages, in its canonical form it does not agree with our analysis and some of the experimental data obtained by the other authors. For the spool-like packaging the association between DNA segmentation and the faces of icosahedron should be the most significant, whereas the strongest association appeared to be with the edges. The restrictions related to dsDNA bending would lead to the voids at the top and bottom parts of capsid, that is not observed on cryo-EM micrographs. Therefore, in comparison with the spool-like mode the spherical-like orientation of DNA in the layers would provide the packaging of DNA longer in total length and may be evolutionary more preferable. The revision of spool-like model can be performed in lines suggested by Hall & Schellman (1982), LaMarque et al. (2004), and Commolli et al. (2008). Their argumentation may be understood within the frameworks of the following simplified model. Consider elastic rode filling the spherical cavity through a gateway. The lowest bending deformation of rode would be along the central circles. The next turns will go near-by up to the significant bending deformations will lead to slipping the winding to another central circle. If a sphere is replaced by icosahedron or dodecahedron, the longer and more stable mechanically turns will lie in the plane between opposite edges rather than between opposite faces. This could explain the association between segmentation and edges and also the dependence of dsDNA conformation on the size and shape of capsid (Petrov et al., 2007). The electrostatic interactions DNA-capsid and DNA-DNA in the presence of counterions and hydration effects (Petrov & Harvey, 2011; Šiber et al., 2012) affect significantly this simplified model but the main features should be retained.

The analysis of large-scale quasi-periodic patterns in the genomic sequences of icosahedral viruses reveals that a part of patterns is universally associated with the elements of icosahedral symmetry. The association of G+C quasi-periodic patterns with the edges appeared to be reproducible in such evolutionary remote viruses as STMV, ϕX174, and HK97. The strongest association was observed mainly between quasi-periodic patterns and the edges as well as the vertices. The reasons for such association may be related to the electrostatic interactions between positively charged capsid and negatively charged DNA or RNA. These interactions are known to play significant role during virion assembly and the genome packaging within capsid (see, e.g., Belyi & Muthukumar, 2006; Hagan, 2009; Ting et al., 2011; and references therein). The general solution of Laplace equation within void charged capsid can be presented as a series on the positive powers of radius and spherical harmonics (see, e.g., Landau & Lifschitz, 1997). This means that DNA-capsid interactions are the strongest at the capsid surface. For a homogeneously charged icosahedral or dodecahedral capsid the electric field would be stronger at the edges and vertices. Therefore, the association between DNA and capsid would be preferable in these regions.

The information stored in underlying genomic sequences is multifarious and is related to all stages of virus life cycle: virion assembly, infection of host cell, replication, encoding genes, and regulation of gene expression. We showed that a part of information related to the genome packaging within icosahedral capsids is also reflected in genomic sequences. Taking into account the incomplete correspondence between genome packaging mode and icosahedral symmetry and the difficulties with the experimental resolution of genome packaging, the study of large-scale quasi-periodic patterns in genomic sequences proved to be surprisingly insightful. The combined bioinformatic, structural, and modeling analysis of the genome packaging within viral icosahedral capsids is of basic interest and may be helpful in development of antiviral drugs. In particular, the genome editing within the stretches of the clustered, regularly-interspaced, short palindromic repeats with CRISPR/Cas9 technique has been applied to *in vitro* and *in vivo* studies for the cure of human diseases including HIV (Soriano, 2017; Lebbink et al., 2017; and references therein). However, the most of repeating patterns are present in the genome as hidden fuzzy repeats. Their involvement in various viral regulatory mechanisms makes them also the promising therapeutic targets. The artificial RNAs with proper AS and PS may be used for the development of virus-like particles for the medical applications (see, e.g., Patel et al., 2017; and references therein).